\documentstyle[12pt]{article}
\input psfig.sty 
\begin{document}
\font\rm=cmr12 
\font\tenrm=cmr10 
\font\tensl=cmsl10
\newcommand{\be}{\begin{eqnarray}} 
\newcommand{\ee}{\end{eqnarray}}
\Huge{\noindent{Istituto\\Nazionale\\Fisica\\Nucleare}}

\vspace{-3.9cm}

\Large{\rightline{Sezione SANIT\`{A}}}
\normalsize{}
\rightline{Istituto Superiore di Sanit\`{a}}
\rightline{Viale Regina Elena 299}
\rightline{I-00161 Roma, Italy}

\vspace{0.65cm}

\rightline{INFN-ISS 96/11}
\rightline{December 1996}

\vspace{1.5cm}

\begin{center}
\LARGE {\bf Neutron electromagnetic form
factors and \\inclusive scattering of polarized electrons\\ by
polarized $^{3}$He and $^{3}$H targets}

\vspace{1cm}

\large{ A. Kievsky$^{1}$, E. Pace$^{2}$, G. Salm\`{e}$^{3}$,
M. Viviani$^{1}$}
\\ [0.4cm] 
$^1$ {\tensl
INFN, Sezione di Pisa, 56010 S.Piero a Grado, Pisa, Italy} 
\\[0.4cm]
$^2${\tensl Dipartimento di Fisica, Universit\`a di Roma "Tor
Vergata", and INFN, 
\\ Sezione Tor Vergata, Via della Ricerca
Scientifica 1, I-00133 Roma, Italy} 
\\[0.4cm] 
$^3${\tensl INFN,
Sezione Sanit\`a, Viale Regina Elena 299, I-00161 Roma, Italy}

\end{center}

\vspace{0.5cm}

\begin{abstract}
\bigskip

\noindent \tenrm{The electromagnetic inclusive responses of polarized
$^{3}$He and $^{3}$H are thoroughly investigated at the quasielastic peak
for squared momentum transfers up to $2~(GeV/c)^2$, within the plane wave
impulse approximation.  Great emphasys is put on the effects in the
bound-state due to different two- and three-body nuclear forces, and to the
Coulomb interaction as well.  A careful analysis of the polarized responses
allows to select possible experiments for minimizing the model dependence
in the extraction of the neutron electromagnetic form factors.  In
particular, the relevant role played by the proton in the
transverse-longitudinal response of polarized $^{3}$He, at low momentum
transfer, can be utilized for obtaining valuable information on the proton
contribution to the total polarized response and eventually on the neutron
charge form factor.}

\end{abstract}

\rm

 \vspace{0.7cm}


\hrule width5cm
\vspace{.2cm}
\noindent To appear in {\bf Phys. Rev. C}

\newpage

\noindent{\bf 1.  Introduction} \bigskip

\pagestyle{plain}

\indent The relevance of an accurate knowledge of nucleon electromagnetic
(em) form factors is well known, and this fact has motivated an impressive
amount of experimental work for investigating these observables.  A
particularly difficult problem is represented by the extraction of the
neutron em form factors, since free neutron targets do not exist in nature,
and therefore one is forced to consider nucleon bound systems, such as
deuteron (see e.g.  \cite{H2}) or polarized $^3$He \cite{BW}.  In the
latter case, within a naive model, with only a symmetric S-wave component
in the bound-state, the two protons have opposite spins and therefore one
should expect that the em polarized response of $^{3}$He essentially is the
neutron one.  Unfortunately, the presence of components with other
symmetries, e.g.  the S'-wave, and higher angular momenta, e.g.  the
D-wave, prevents the application of the naive picture and therefore also
the polarized $^3$He target is plagued by a non negligible proton
contribution.  In spite of this, many experimental efforts
\cite{Jon,Thom,Gao,Han} have been carried out with the aim of measuring the
inclusive response of polarized $^3$He at the quasielastic (qe) peak.
Though such measurements were affected by sizable statistical and
systematic uncertainties, a first estimate of the neutron magnetic form
factor at $Q^2\approx 0.2~(GeV/c)^2$ \cite{Gao} was extracted from the
transverse polarized response (with a $\approx~ 20 \%$ accuracy), while no
information on the charge form factor could be reliably obtained
\cite{Han}, due to the large proton contribution in the
transverse-longitudinal polarized response.  In order to improve the
accuracy, in particular for the transverse polarized response, a new
generation of inclusive experiments to be performed in the near future at
TJNAF has been planned (see e.g.  \cite{TJNAF}).  On the theoretical side,
the inclusive polarized responses have been analyzed within the {\em plane
wave impulse approximation} (PWIA), by calculating the spin-dependent
spectral function of $^3$He \cite{CPS,Sauer} from realistic wave functions
of the three-nucleon system, pointing out the unpleasant presence of the
proton contribution, particularly in the transverse-longitudinal response
investigated in the experiments of Refs.  \cite{Jon,Thom,Han}.
Nevertheless, a suitable choice of the polarization angle can minimize the
proton contribution, as discussed in detail in \cite{CPS}.  It is worth
noting that i) the inclusion of the final state interaction (FSI), between
the knocked out nucleon and the spectator pair, ii) the meson exchange
currents, iii) the relativistic corrections and iv) the presence of the
$\Delta$ in the bound-state, are still open problems for the inclusive
polarized responses of the three-nucleon system.  As far as the first topic
is concerned, that only for unpolarized cross sections refined calculations
including FSI are available \cite{Gol,Mar}, exhibiting large FSI effects at
very low momentum transfer ($Q^2~<~0.1~(GeV/c)^2$), while at
$Q^2~>~0.25~(GeV/c)^2$ PWIA calculations are able to give a good
description of the experimental data at the qe peak, both for the
unpolarized longitudinal and transverse response functions \cite{Ben,Mei}.

\indent Our aim is to study, within the PWIA, em inclusive polarized
responses of both $^3$He and $^3$H, at the qe peak, for a wide range of
momentum transfer, in order to explore the model dependence due to the
initial state interaction (ISI) (see also \cite{Han} for the particular
case of the corresponding experiment).  We have obtained the spin-dependent
spectral function from bound-state wave functions, calculated using the
Pair correlated Hyperspherical-Harmonic (PHH) basis \cite{KVR} and
different realistic two- and three-body nuclear forces; moreover, we have
also considered the Coulomb interaction in $^3$He.  After a detailed
analysis of the effects of two- and three-body interactions on the
polarized responses, we have singled out possible experiments for
minimizing the model dependence due to both the nuclear structure and the
presence of the proton contribution, when neutron form factors are
extracted.  First of all, we propose a measurement of the
transverse-longitudinal polarized response of $^3$He at low momentum
transfer, in the range $0.1 \le Q^2 \le 0.3~ (GeV/c)^2$, since one could
obtain valuable information on the proton contribution to this response,
just taking advantage of the proton predominance in this kinematical
region.  Moreover, a measurement of the polarization angle where the proton
contribution to the polarized cross section is vanishining, could give the
ratio between the proton contributions to the transverse response and the
transverse-longitudinal one; it will be pointed out that an estimate of
such a polarization angle can be obtained through a measurement of protons
emitted along the direction of the three-momentum transfer.  Finally,
future experiments for measuring the em response of polarized $^3$H could
close the chain for obtaining an almost model independent extraction of
both electric and magnetic neutron form factors.

\indent The paper is organized as follows:  in Sect.  2, the inclusive
cross section of polarized electrons by a polarized three-nucleon target
will be reported, as well as the nuclear polarized response functions
obtained within the PWIA; in Sect.  3, the three-nucleon ground-state and
the spin-dependent spectral functions we have adopted will be illustrated
in detail; in Sect.  4, the comparison between our results, calculated by
using different realistic two- and three-body interactions, and the most
recent experimental values for $^3$He asymmetries \cite{Gao,Han} will be
presented, and em polarized responses at qe peak, for both $^3$He and
$^3$H, will be investigated in detail up to $Q^2=2~(GeV/c)^2$; in Sect.  5,
possible experiments for minimizing the model dependence in the extraction
of neutron em form factors will be suggested; in Sect.  6 conclusions will
be drawn.

\bigskip

\noindent{\bf 2.  The inclusive cross section and the em responses of a J =
1/2 nucleus}

\bigskip

\indent In this Section the PWIA formalism we have adopted will be
presented, with a particular emphasysis on the em responses in terms of the
spin-dependent spectral function of the three-body system.

\indent First of all, the inclusive cross section of a longitudinally
polarized electron with helicity $h=\pm1$ from a $J=1/2$ nucleus is
reported for the sake of completeness.  After contracting the em tensors
for the electron and for the nuclear target, A, the cross section can be
cast in terms of the unpolarized ($R^A_{L}$, $R^A_{T}$) and polarized
($R^A_{T'}$, $R^A_{TL'}$) nuclear response functions as follows (for a
detailed discussion see \cite{CPS}, cf.  also \cite{Sauer,Don})

\be \frac{d^2\hbox{$\large\sigma$}(h)}{d\Omega_2 d\nu}~=~
\hbox{$\large\Sigma$}\;+\;h\;\hbox{$\large\Delta$} \label{eq14} \ee with

\be \hbox{$\large\Sigma$}~=~ \hbox{$\large\sigma$}_{Mott} \left [
 \left({Q^2 \over |\vec{q}|^2}\right )^2R^A_{L}(Q^2,\nu)\;+\;\left({Q^2
 \over 2|\vec{q}|^2}+tan^2{\theta_{e}\over
 2}\right)~R^A_{T}(Q^2,\nu)\right] \label{Sigma} \ee \be
 \hbox{$\large\Delta$}~=~
 -\hbox{$\large\sigma$}_{Mott}~tan\frac{\theta_{e}}{2}
 \left\{cos\theta^*~R^A_{T'}(Q^2,\nu)~\left[{Q^2 \over |\vec{q}|^2} +
 tan^2\frac{\theta_{e}}{2}\right]^{1/2}~+ \right.  \nonumber \\ \left.
 -~{Q^2 \over |\vec{q}|^2~\sqrt{2}}~sin\theta^*
 cos\phi^*~R^A_{TL'}(Q^2,\nu)\right \} \label{Delta} \medskip \ee where
 $\theta_{e}$ is the scattering angle; $\theta^* $ and $\phi^*$ are the
 azimuthal and polar angles of the target polarization vector $\vec{S_A}$,
 with respect to the direction of the three-momentum transfer $\vec{q}$;
 $Q^2=|\vec{q}|^2 - \nu^2$.  In our analysis, in addition to $\theta^* $,
 another polarization angle, $\beta$, will be used, defined with respect to
 the direction of the incoming electron beam, i.e.  $cos\beta\equiv
 \vec{S_A}\cdot \vec{k}_1/|\vec{k}_1| $, where $\vec k_{1(2)}$ is the
 three-momentum of the initial (final) electron.  The relation between the
 polarization angles we have considered is \be cos\theta^*= {\epsilon_1~
 cos \beta ~-~\epsilon_2~ (sin \beta~sin\theta_e~cos \phi ~+~cos \beta~cos
 \theta_e) \over|\vec{q}|} \nonumber \\
 sin\theta^*cos\phi^*={|\vec{q}|cos\beta~-(\epsilon_1~-~\epsilon_2
 cos\theta_e) cos\theta^* \over \epsilon_2 sin\theta_e} \label{ang}
 \medskip \ee where $\phi$ is the polar angle with respect to $
 \vec{k}_1/|\vec{k}_1|$.  In the case of coplanar kinematics (i.e.  $\phi
 =0^o~,~ 180^o$) one can have only $\phi^*=0^o$ or $ 180^o$.

Following \cite{CPS}, the nuclear response functions can be expressed in
terms of a 2x2 matrix ${\bf\hat{P}}^{N}_{\cal{M}}(\vec{p},E)$ representing
the spin dependent spectral function of a nucleon, $N$, inside a nucleus
with component of the total angular momentum along the polarization
$\vec{S}_A$ equal to $\cal M$.  The elements of the matrix
${\bf\hat{P}}^{N}_{\cal{M}}(\vec{p},E)$ are given by \be P_{\sigma,
\sigma',\cal{M}}^{N} ({\vec{p},E})=\sum\nolimits\limits_{{f}_{(A-1)}}
~_{N}\langle{\vec{p},\sigma;\psi }_{f_{(A-1)}} |{\psi
}_{J\cal{M}}\rangle~\langle{\psi }_{J\cal{M}}|{\psi
}_{f_{(A-1)}};\vec{p},\sigma '\rangle _{N}~ \delta
(E-{E}_{f_{(A-1)}}+{E}_{A}) \label{eq9} \medskip \ee where $|{\psi
}_{J\cal{M}}\rangle$ is the ground state of the target nucleus with
polarization $\vec{S}_A$, $|{\psi }_{f_{(A-1)}}\rangle$ an eigenstate of
the spectator system with quantum numbers $f$ and interacting through the
{\em same interaction} of the target nucleus, $|\vec{p},\sigma\rangle_N$ a
plane wave describing the nucleon N with spin component, along the z-axis,
equal to $\sigma$; $E$ is the missing energy.  In a more compact form, for
$J=1/2$, ${\bf\hat{P}}^{N}_{\cal{M}}(\vec{p},E)$ can be written as \be
{\bf\hat{P}}^{N}_{\cal{M}}(\vec{p},E)={1\over 2}\left
\{B_{0,{\cal{M}}}^{N}(|\vec{p}|,E)~+~\vec{\sigma} \cdot \left [
\vec{S}_A~B_{1,\cal{M}}^{N}(|\vec{p}|,E)~+~\hat{p}~(\hat{p} \cdot
\vec{S}_A)~ B_{2,\cal{M}}^{N}(|\vec{p}|,E)\right] \right \} \label{eq9.1}
\medskip \ee where the function $B_{0,\cal{M}}^{N}(|\vec{p}|,E)$ is the
trace of ${\bf\hat{P}}^{N}_{\cal{M}}(\vec{p},E)$ and yields the usual
unpolarized spectral function, while $B_{1,\cal{M}}^{N}(|\vec{p}|,E)$ and
$B_{2,\cal{M}}^{N}(|\vec{p}|,E)$ describe the spin structure of the
probability distribution of finding a nucleon with a given momentum,
missing energy and polarization.  As a matter of fact the effective
polarization of a nucleon is given by \be {\cal{P}}^N= \langle{\psi }_{J {1
\over 2}}|\vec{S}_A \cdot \vec{\sigma}|{\psi }_{J {1 \over 2}}\rangle~=~
\int dE~\int d\vec{p}~{\bf Tr} \left [\vec{S}_A \cdot \vec{\sigma}~
{\bf\hat{P}}^{N}_{{1 \over 2}}(\vec{p},E) \right ] ~=\nonumber \\
=~4\pi~\int dE~\int p^2~d{p}~ \left (B_{1,{1 \over 2}}^N(|\vec{p}|,E)~+~
{B_{2,{1 \over 2}}^N(|\vec{p}|,E) \over 3} \right ) \label{pol} \medskip
\ee In Sect.  3, where more details will be given on the spin-dependent
spectral function, the values of ${\cal{P}}^N$ in $^3$He will be also
presented for the different interactions we have used, but we can
anticipate that ${\cal{P}}^n$ is about $86-87~\% $, in line with the
analysis of the world calculations of Ref.  \cite{Friar}.  Explicit
expressions for the functions $B_{0(1,2),\cal{M}}$ can be found in
\cite{CPS}.

\indent The evaluation of the response functions $R^A_{L}$, $R^A_{T}$,
$R^A_{T'}$ and $R^A_{TL'}$ requires an off-energy-shell em nucleon tensor
in the PWIA convolution formulas.  In the literature different recipes for
the off-energy-shell em nucleon current have been proposed (see e.g.
\cite{DeF,Pol}).  In particular, in the widely adopted prescriptions CC1
and CC2 of \cite{DeF} the longitudinal component of the em current is
eliminated in favour of the charge component in order to restore the
current conservation (Coulomb gauge).  In the actual calculations we have
adopted such an approximation, since we have checked that in PWIA it allows
for a good description of the accurate experimental data for the
unpolarized $^3$He longitudinal and transverse responses at $Q^2~>~
0.25~(GeV/c)^2$ (see also \cite{Ben,Mei}), while the approximation, based
on the elimination of the charge component in favour of the longitudinal
one (Weyl gauge), underestimates the longitudinal response at the top of
the qe peak (it should be pointed out that the possible FSI effects are
expected to further reduce the response at the top of the qe peak
\cite{Gol}, worsening the comparison with the data).  It turns out that in
the Landau gauge ( i.e.  $J^{\mu}~\rightarrow~ J^{\mu}+q^{\mu} J \cdot
q/Q^2$) the responses are essentially the same as in the Coulomb one.  In
the following only the expressions corresponding to the CC1 prescription
are reported, since the numerical results obtained by using CC2 slightly
differ (see Sect.  4).  After a lengthy algebra (cf.  \cite{CPS} for the
on-energy-shell case) one gets (in the following, $p~\equiv~|\vec{p}|$ and
$q~\equiv~|\vec{q}|$ )

\be R^A_{L} \left(Q^2,\nu \right )= {\pi \over 2}
 \sum\nolimits\limits_{N=p,n} {\cal {N}}_A^{N} \int \limits_{E_{min}}
 \limits^{E_{max}(Q^2,\nu)}dE\int
 \limits_{p_{min}(Q^2,\nu,E)}\limits^{p_{max}(Q^2,\nu,E)}{p\over{qE_p}}~dp~
 B_0^{N}\!\left( p,E\right) ~~~~~~~~~~~~~~~ \nonumber \\ \left \{ \rule
 {0.cm} {.8cm} \left(2E_p+\bar{\nu} \right )^2 \left ( \left(
 {F_{1}^{N}}\!\left({Q}^{2} \right)\right ) ^2+{\bar \tau} \left
 ({F_{2}^{N}}\!\left({Q}^{2} \right) \right ) ^2\right) -q^2 \left(
 F_{1}^{N}\!\left({Q}^{2} \right)+F_{2}^{N}\!\left({Q}^{2} \right)
 \right)^2 \rule {0.cm} {.8cm} \right \} \label{RL} \medskip \ee

\be R^A_{T} \left(Q^2,\nu \right )= \pi \sum\nolimits\limits_{N=p,n} {\cal
 {N}}_A^{N} \int \limits_{E_{min}} \limits^{E_{max}(Q^2,\nu)}dE\int
 \limits_{p_{min}(Q^2,\nu,E)}\limits^{p_{max}(Q^2,\nu,E)}{p\over{qE_p}}~dp~
 B_0^{N}\!\left( p,E\right)~~~~~~~~~~~~~~~ \nonumber \\ \left \{ \rule
 {0.cm} {.8cm} {\bar Q}^2 \left( F_{1}^{N}\!\left({Q}^{2}
 \right)+F_{2}^{N}\!\left({Q}^{2} \right) \right)^2 +2 p^2 sin^2\alpha
 \left( \left({F_{1}^{N}}\!\left({Q}^{2} \right)\right)^2+{\bar \tau} \left
 (F_{2}^{N}\!\left({Q}^{2} \right) \right)^2\right)\rule {0.cm} {.8cm}
 \right \} \label{RT} \medskip \ee

\be R^A_{T'} \left(Q^2,\nu \right )=2 \pi \sum\nolimits\limits_{N=p,n}
 {\cal {N}}_A^{N}~\left( F_{1}^{N}\!\left({Q}^{2}
 \right)+F_{2}^{N}\!\left({Q}^{2} \right) \right) \int \limits_{E_{min}}
 \limits^{E_{max}(Q^2,\nu)}dE\int
 \limits_{p_{min}(Q^2,\nu,E)}\limits^{p_{max}(Q^2,\nu,E)}{p\over{qE_p}}~dp
 \nonumber \\ \left \{ \rule {0.cm} {.8cm}
 p~cos\alpha~\left(B_1^{N}\!\left( p,E\right)~+B_2^{N}\!\left( p,E \right)
 \right) \left[ \left ( F_{1}^{N}\!\left({Q}^{2}
 \right)-F_{2}^{N}\!\left({Q}^{2} \right){{\bar Q}^2 \over {2 M {\bar
 \nu}}} \right) {{\bar \nu} p~cos\alpha \over{M+E_p}} -
 ~F_{1}^{N}\!\left({Q}^{2} \right) q \right ]~+~ \right .  \nonumber \\
 \left .  +~ \left( B_1^{N}\!\left( p,E \right)~+B_2^{N}\!\left( p,E
 \right)cos^2 \alpha \right)~ M {\bar{\nu}} ~\left(F_{1}^{N}\!\left({Q}^{2}
 \right)+F_{2}^{N}\!\left({Q}^{2}\right){{\bar Q}^2 E_p \over {2 M^2 {\bar
 \nu}}} \right)\rule {0.cm} {.8cm} \right \} ~~~~~~ \label{RT'} \medskip
 \ee

\be R^A_{TL'} \left(Q^2,\nu\right )=- \sqrt{2}~ q~ 2\pi
 \sum\nolimits\limits_{N=p,n} {\cal {N}}_A^{N}~\left(
 F_{1}^{N}\!\left({Q}^{2} \right)+F_{2}^{N}\!\left({Q}^{2} \right)\right)
 \int \limits_{E_{min}} \limits^{E_{max}(Q^2,\nu)}dE\int
 \limits_{p_{min}(Q^2,\nu,E)}\limits^{p_{max}(Q^2,\nu,E)}{p\over{qE_p}}~dp
 \nonumber \\ \left \{ \rule {0.cm} {.8cm}\left(B_1^{N}\!\left(
 p,E\right)~+B_2^{N}\!\left( p,E \right) \right) \left
 (F_{1}^{N}\!\left({Q}^{2} \right)+F_{2}^{N}\!\left({Q}^{2} \right){{\bar
 {\nu}} \over {2 M }} \right){p^2~sin^2\alpha \over{2(M+E_p)}} ~+~ \right.
 \nonumber \\ \left.  + ~F_{2}^{N}\!\left({Q}^{2}~ \right){p~cos\alpha
 \over 2 M}~qB_1^{N}\!\left( p,E\right) + \right.  \nonumber \\ \left.
 +~\left(B_1^{N}\!\left( p,E\right)~+{B_2^{N}\!\left( p,E \right) \over 2
 }sin^2 \alpha \right)
 M~\left(F_{1}^{N}\!\left({Q}^{2}\right)-F_{2}^{N}\!\left({Q}^{2}\right){{\bar
 \nu} E_p \over {2 M^2 }} \right) \rule {0.cm} {.8cm}\right \}~
 \label{RTL'} \medskip \ee where $B^N_{0(1,2)} \equiv B^N_{0(1,2),{1 \over
 2}}$ are the functions corresponding to the target nucleus,
 $~F_{1(2)}^{N}$ is the Dirac (Pauli) nucleon form factor,
 ${\cal{N}}^{p(n)}_A$ the number of proton (neutron) in the nucleus A,
 $E_p=\sqrt{M^2+p^2}$, ${\bar \nu}=\sqrt{M^2+p^2+q^2+2pqcos \alpha}~ -
 ~E_p$, ${\bar Q}^2=|\vec{q}|^2-{\bar \nu}^2$ and $\bar{\tau}={\bar
 Q}^2/4M^2$.

In Eqs.  (\ref{RL}) - (\ref{RTL'}) the integration limits and $cos \alpha$
are determined, as usual, by energy conservation \cite{CPS2}.

\bigskip

\noindent{\bf 3.  The three-body ground-state in the
hyperspherical-harmonic method and the spin-dependent spectral function}

\bigskip

The spin-dependent spectral functions of $^3$He and $^3$H, Eqs.(\ref{eq9})
and (\ref{eq9.1}), have been calculated from three-body wave functions
obtained using the PHH expansion.  This technique represents a very
efficient and accurate method for describing the system, and explicitely
includes pair correlation functions in order to take into account the short
range repulsion of the nucleon-nucleon (NN) interaction.  In the following
the main features of the method \cite{KVR} will be briefly recalled.

The three-nucleon wave function with total angular momentum $J{\cal M}$ and
total isospin $TT_z$ can be written, in the $LS$ coupling scheme, as \be
\psi^{TT_z}_{J{\cal M}}&=&\sum_{i=1,3}\left\{ \sum_{\alpha=1}^{N_c}
\Phi_\alpha(x_i,y_i) {\cal Y}_\alpha (jk,i) \right\} \label{eq2.0} \medskip
\ee with \be {\cal Y}_\alpha (jk,i) &=& \Bigl\{\bigl[ Y_{\ell_\alpha}(\hat
x_i) Y_{L_\alpha}(\hat y_i) \bigr ]_{\Lambda_\alpha} \bigl [ s_\alpha^{jk}
s_\alpha^i \bigr ]_{S_\alpha} \Bigr \}_{J {\cal M}} \; \bigl [
t_\alpha^{jk} t_\alpha^i \bigr ]_{T T_z} \label{eq2.1} \medskip \ee where
$x_i$, $y_i$ are the moduli of the Jacobi coordinates, $\alpha$ runs over
the three--body channels included in the partial wave decomposition of the
wave function and $(jk,i)$ are cyclic permutations of $(12,3)$.  The total
number of channels considered is $N_c$.  The two--dimensional amplitudes
$\Phi_\alpha(x_i,y_i)$ are expanded in terms of the PHH basis
\begin{equation}
\Phi_\alpha(x_i,y_i)=\rho^{\ell_\alpha+L_\alpha}f_\alpha(x_i)
\Bigl[\sum_{K=K_0}^{K_\alpha} u^\alpha_K(\rho)~~{}^{(2)}P^{\ell_\alpha
+L_\alpha}_K(\phi_i)\Bigr] \label{eq2.2} \end{equation} where $x_i =
\rho\cos(\phi_i)$ and $ y_i = \rho\sin(\phi_i)$ are the hyperspherical
variables, $K_0=\ell_\alpha + L_\alpha$ and
${}^{(2)}P^{\ell_\alpha+L_\alpha}_K(\phi_i)$ is an hyperspherical
polynomial \cite{Fabr}.  The pair correlation functions $f_\alpha(x_i)$ are
introduced in order to accelerate the convergence of the expansion and are
determined by the NN potential as explained in \cite{KVR}.  The unknown
quantities in Eq.(\ref{eq2.2}) are the hyperradial functions
$u^\alpha_K(\rho)$, which are obtained through the Rayleigh-Ritz
variational principle.

Convergence and accuracy reached by increasing the number of terms in the
expansions (\ref{eq2.0}) and (\ref{eq2.2}) are discussed in \cite{KVR} for
realistic NN potentials with and without three-nucleon interaction (TNI)
terms included in the Hamiltonian of the system.  Typically with $N_c=18$
and about 80 hyperradial functions an accuracy of the order of a few keV is
obtained for the binding energy of the system.  In the calculations of the
spin-dependent spectral function we have used three-body wave functions
corresponding to a variety of NN potentials and TNI's.  In particular we
have considered:  i) NN interactions such as the Argonne Av14 interaction
\cite{Arg} and the Reid soft-core \cite{RSC} RSCv8 potential (that
represents the local version of the original one with eight operators
only), and ii) the Brazil three-body force \cite{BR} (BR) and the
Tucson-Melbourne one \cite{TM} (TM).  When the TNI have been taken into
account, the cut-off parameter was adjusted in order to give the triton
experimental binding energy $B(^3{\rm H})=8.48$ MeV.  It should be
emphasyzed that for the $^3$He ground-state the Coulomb interaction between
the two protons has been included, and this possibility represents a
typical feature of the variational approaches.  In Table I, binding
energies and probabilities of symmetric-S, S', D, and P components of $^3$H
and $^3$He wave functions, corresponding to the interactions we have
adopted, are listed.  The interactions we considered have sizable
differences in their structures (e.g.  the RSCv8 interaction has a
Yukawa-type repulsion at short distances, whereas the Av14 interaction
remains finite), and these differences are reflected in the corresponding
three-body wave functions.  In particular the percentages of the small
components are affected by the choice of the potential, as illustrated,
e.g., by $P_{S'}$ that varies up to 30$\%$ by changing from RSCv8
interaction to Av14 interaction + three-body forces.

In order to calculate $_N\langle{\vec{p},\sigma;\psi }_{f_{(A-1)}} |{\psi
}^{T,T_z}_{J,\mu}\rangle$, see Eq.(\ref{eq9}), we Fourier-transformed the
following overlap \be
G^{T_z,\mu}_{f,\sigma,\tau}(\vec{y})=\langle{\chi^{1/2}_{\sigma}~
\xi^{1/2}_{\tau }\vec{y};\psi}_{f} |{\psi }^{1/2,T_z}_{1/2,\mu} \rangle =
\nonumber \\ =~\langle {1 \over 2} \tau T_{12} \tau_{12} |{1 \over 2} T_z
\rangle ~\sum_{\ell~m~j~m_j} \langle J_{12} m_{12} {1 \over 2} \sigma |jm_j
\rangle \langle \ell m jm_j|{1 \over 2} \mu \rangle ~g^f_{\ell
j}(y)~Y_{\ell m}(\hat y) \label{eq2.4} \medskip \ee where $f\equiv [J_{12},
~m_{12}, ~S_{12},~T_{12},~\tau_{12},~\lambda,~E_{12}]$ (cf.  \cite{CPS1})
represents the quantum numbers of the two-body wave function
$|\psi_{f}({\vec{ x}})\rangle$, corresponding to the particles (1,2) that
interact with the same potential as in the three-body hamiltonian;
$\chi^{1/2}_{\sigma}$ and $\xi^{1/2}_{\tau}$ are the spin and isospin
functions of the third particle.  The functions $g^f_{\ell j}(y)$ are the
overlaps between the three-body ground-state and the two-body wave function
coupled to the third particle spin, isospin, and orbital momentum
eigenfunctions.  Once the functions $G^{T_z,\mu}_{f,\sigma,\tau}$ have been
obtained, the evaluation of the functions $B^{N}_{0(1,2),\cal{M}}$ is
performed following the expressions of Ref.\cite{CPS}.

Before presenting the actual results of these functions for the case of
$^3$He, it is interesting to consider the effective polarization of the
nucleon, ${\cal{P}}^N$, Eq.(\ref{pol}), that represents an integral
property of the three-body wave function.  \begin{table}

\begin{quote} {\tenrm Table I:  Binding energies and probabilities of
symmetric-S, S', D and P waves for $^3$H and $^3$He obtained with different
ISI (see text for the legend).}  \end{quote} \rm \begin{center}
\begin{tabular}{|l|c||c||c||c||c|} \hline \hline
\multicolumn{6}{|c|}{$^3$H} \\ \hline ISI & $B$(MeV) & $P_S$(\%)&
$P_{S'}$(\%) & $P_D$(\%) &$P_P$(\%) \\ \hline Av14 & 7.683 &89.831 & 1.126&
8.967& 0.076 \\ \hline RSCv8 & 7.600 &88.920 & 1.342& 9.654& 0.084 \\
\hline Av14 + BR & 8.485 &89.388 & 0.928 & 9.544 & 0.140 \\ \hline Av14 +
TM & 8.485 &89.645 & 0.933 & 9.261 & 0.161 \\ \hline \hline
\multicolumn{6}{|c|}{$^3$He} \\ \hline ISI & $B$(MeV) & $P_S$(\%)&
$P_{S'}$(\%) & $P_D$(\%) &$P_P$(\%) \\ \hline Av14 & 7.032 &89.680 & 1.314&
8.931& 0.075 \\ \hline RSCv8 & 6.958 &88.757& 1.550& 9.610& 0.083 \\ \hline
Av14 + BR & 7.809 &89.276 & 1.077 & 9.509 & 0.138 \\ \hline Av14 + TM &
7.809 &89.526 & 1.083 & 9.233 & 0.158 \\ \hline \end{tabular} \end {center}
\end{table} In Table II the values of ${\cal{P}}^N$ for $^3$He are listed
for the different interactions we have used (the corresponding values for
$^3$H can be obtained by isospin symmetry once the Coulomb interaction is
disregarded).  In particular, for the sake of completeness, we have also
reported the effective polarization corresponding to the case when the
Coulomb interaction is switched off in $^3$He.  As pointed out in
\cite{Friar} the differences between ${\cal{P}}^N$ values can be traced
back to the differences in the probabilities of S' and D waves (see Table
I).  It is worth noting that Coulomb potential and TNI effects are small
and become important, in percentage, for the proton effective polarization,
since this quantity is small by itself.

\begin{table}

\begin{quote} {\tenrm Table II.:  The values of the nucleon effective
polarization in $^3$He, obtained with different ISI (see text for the
legend).}  \end{quote} \rm \begin{center} \begin{tabular}{|l|c||c|} \hline
~~~ISI~~~~~~~~~ & ~~${\cal{P}}_n~\%$~~ & ~~${\cal{P}}_p~\%$ \\ \hline Av14
& 87.37& -2.57 \\ \hline Av14+Cou & 87.30& -2.49 \\ \hline Av14+Cou+BR &
86.65& -2.77 \\ \hline Av14+Cou+TM & 87.01& -2.68 \\ \hline RSCv8 & 86.12&
-2.67 \\ \hline RSCv8+Cou & 86.19& -2.76 \\ \hline \end{tabular} \end
{center} \end{table}

In Figs.  1 and 2, the unpolarized spectral function $ B_{0,{1 \over
2}}^{p(n)}(|\vec{p}|,E)$ and the functions $|B_{1,{1 \over
2}}^{p(n)}(|\vec{p}|,E)|~$ and $|B_{2,{1 \over 2}}^{p(n)}(|\vec{p}|,E)|$
obtained from the $^3$He wave function corresponding to the Av14 potential
model {\em plus} the Coulomb interaction, are shown (in the case of the
proton the curve corresponding to a spectator deuteron is not presented).
It has to be pointed out that the introduction of three-body forces does
not change sizably the overall behaviour, and produces only small
differences in the values for any given (E,p); these differences could be
important in the calculation of tiny effects, such as the cross section of
exclusive reactions where the proton is detected, since large cancellations
between contributions produced by a spectator pair in the deuteron state
and in the continuum occur.  The spin-dependent spectral function was
already calculated, without Coulomb interaction:  i) in Ref.  \cite{CPS},
adopting a variational three-body wave function corresponding to the Reid
soft-core interaction \cite{RSC}; and ii) in Ref.\cite{Sauer}, using a
Faddeev wave function and the Paris potential \cite{Paris}.  The comparison
between our results of Figs.  1 and 2, obtained without three-body forces
but with the Coulomb interaction, and the corresponding ones of \cite{CPS},
obtained with a different two-body nuclear force and no Coulomb
interaction, illustrates the sensitivity to the choice of the two-body part
of the interaction, showing that the gross features remain unchanged.
Relevant changes can be found only in $|B_{1,{1 \over
2}}^{p(n)}(|\vec{p}|,E)|~$ and $|B_{2,{1 \over 2}}^{p(n)}(|\vec{p}|,E)|$
(cf.  also Table II).  Moreover, a smoother behaviour is observed in the
present spectral functions, due to the improved accuracy in the description
of the bound-state wave functions.

\bigskip

\noindent{\bf 4.  The em polarized responses in $^3$He and $^3$H}

\bigskip

\indent The nuclear responses, for both $^3$He and $^3$H, have been
obtained by using the spin-dependent spectral functions calculated from the
variational wave functions described in Sect.  3.  The results
corresponding to CC1 and CC2 prescriptions for the off-energy-shell em
nucleon tensor are essentially the same for $R_{TL'}$, while, in the whole
range $0.1 \leq Q^2 \leq 2 (GeV/c)^2$ we have investigated, the differences
for the proton and neutron contributions to $R_{T'}$ are in general less
than 2$\%$, except a 3$\%$ variation for the negligible neutron
contribution in $^3$H.  Therefore in the following we report only the
results obtained by the prescription CC1, i.e.  by using Eqs.
(\ref{RL})-(\ref{RTL'}).

\indent Before analyzing in detail the em polarized responses, let us
compare our predictions for the asymmetry of $^3$He, defined by \be
A={\hbox{$\large\Delta$} \over \hbox{$\large\Sigma$}} \label{Asy} \medskip
\ee (cf.  Eqs.  (\ref{Sigma}) and (\ref{Delta})), with the most recent data
from inclusive experiments that should be proportional, at qe peak, to
$R^{^3He}_{T'}$ \cite{Gao} and $R^{^3He}_{TL'}$ \cite{Han}.  In Figs.  3
and 4, the results obtained by using different ISI are shown.  In
particular, in Figs.  3(a) and 4(a) the theoretical predictions
corresponding to:  i) Av14 potential + Coulomb interaction, ii) RSCv8
potential + Coulomb interaction and iii) Paris potential without Coulomb
interaction, from \cite{Sauer}, are compared with the experimental data; it
is worth noting that for the kinematics of Ref.  \cite{Gao}, where
$A_{peak}\propto R^{^3He}_{T'}$, the asymmetry does not appreciably change
by varying the two-body ISI (less than 2$\%$ at the qe peak), while for the
kinematics of Ref.  \cite{Han}, where $A_{peak}\propto R^{^3He}_{TL'}$,
there is a little bit higher variation (less than 4$\%$ at the qe peak), in
line with the findings of the theoretical analysis contained in \cite{Han}.
In particular for the latter kinematics, the differences among the
theoretical curves are smaller than the ones found in \cite{Han}, where the
original RSC interaction without Coulomb potential was considered.
Differences are even smaller if we consider the effects of three-body
forces, as shown in Figs.  3(b) and 4(b).  These results are quite
reasonable in view of the tiny ISI effects on the effective polarizations
shown in Table II.  One can conclude that at least for low values of
momentum transfer ($Q^2\le 0.2 ~(GeV/c)^2$), changes in the asymmetry due
to ISI are small at the qe peak and therefore the model dependence seems to
be under control, since it amounts to 3-4$\%$, at most.  In Figs.  3 and 4
the neutron contribution to the total responses is shown separately.  From
a comparison between such a contribution and the total responses one can
see that the presence of the proton, in particular for $R^{^3He}_{TL'}$,
represents the major obstacle in the extraction of neutron em form factors,
as already pointed out in \cite{Han,CPS,Sauer}.

\indent In what follows we will carry out a systematic investigation of the
model dependence, due to ISI and proton contribution, in the extraction of
neutron form factors for $Q^2$ up to 2 $(GeV/c)^2$, i.e.  for a kinematical
region relevant for TJNAF.  Moreover, we will show the corresponding
results for $^3$H in order to gain some insight on the possibility to
reduce the model dependence by taking advantage of the isospin symmetry,
partially broken by the Coulomb interaction.

\indent The analysis of the polarized responses at qe peak is motivated by
the expectation of a clear disentangling of the nuclear structure from the
nucleon form factors.  For $^3$He and $^3$H, even if the FSI are present,
Eqs.  (\ref{RT'}) and (\ref{RTL'}) can be written as follows \be
R^{^3He}_{T'} \left(Q^2,\nu \right)= {Q^2 \over 2 q M} \left [2~\left(
G_{M}^{p}\!\left({Q}^{2}\right)\right )^2 {\cal H}^p_{T'}\left(Q^2,\nu
\right)+\left( G_{M}^{n}\!\left({Q}^{2}\right)\right )^2 {\cal
H}^n_{T'}\left(Q^2,\nu \right) \right ] \label{RTHe} \ee

\be R^{^3He}_{TL'} \left(Q^2,\nu \right)= - \sqrt{2} \left [
 2~G_{E}^{p}\!\left({Q}^2\right ) G_{M}^{p}\!\left({Q}^2\right ) {\cal
 H}^p_{TL'}\left(Q^2,\nu \right)+G_{E}^{n}\!\left({Q}^2\right )
 G_{M}^{n}\!\left({Q}^2\right ) {\cal H}^n_{TL'}\left(Q^2,\nu \right)
 \right ] \label{RTLHe} \ee

\be R^{^3H}_{T'} \left(Q^2,\nu \right)= {Q^2 \over 2q M} \left [ \left(
 G_{M}^{p}\!\left({Q}^{2}\right)\right )^2 {\cal T}^p_{T'}\left(Q^2,\nu
 \right)+2~\left( G_{M}^{n}\!\left({Q}^{2}\right)\right )^2 {\cal
 T}^n_{T'}\left(Q^2,\nu \right) \right ] \label{RTH} \ee

\be R^{^3H}_{TL'} \left(Q^2,\nu \right)= - \sqrt{2} \left [
 G_{E}^{p}\!\left({Q}^2\right ) G_{M}^{p}\!\left({Q}^2\right ) {\cal
 T}^p_{TL'}\left(Q^2,\nu \right)+2~G_{E}^{n}\!\left({Q}^2\right )
 G_{M}^{n}\!\left({Q}^2\right ) {\cal T}^n_{TL'}\left(Q^2,\nu \right)
 \right ] \label{RTLH} \ee where the Sachs form factors,
 $G_{M}^{N}=F_{1}^{N}+F_{2}^{N}$ and $G_{E}^{N}=F_{1}^{N}-F_{2}^{N}{Q^2
 \over {4 M^2 }}$, have been introduced.  The functions ${\cal
 H}^N_{T'(TL')}$ and ${\cal T}^N_{T'(TL')}$ in general contain both the
 nuclear structure, which within PWIA is described by $B_{0,1,2}$, and
 ratios of nucleon form factors, and are defined by \be {\cal
 H}^{p(n)}_{T'(TL')} \left(Q^2,\nu
 \right)={R^{^3He,p(n)}_{T'(TL')}\left(Q^2,\nu \right)\over
 K^{p(n)}_{T'(TL')}\left(Q^2 \right)} \nonumber \\ {\cal
 T}^{p(n)}_{T'(TL')} \left(Q^2,\nu
 \right)={R^{^3H,p(n)}_{T'(TL')}\left(Q^2,\nu \right)\over
 K^{p(n)}_{T'(TL')}\left(Q^2\right)}\label{HT} \ee with
 $R^{^3He,p(n)}_{T'(TL')}$ and $R^{^3H,p(n)}_{T'(TL')}$ the proton
 (neutron) contribution to the total responses, $ K^{p(n)}_{T'}= Q^2
 {\cal{N}}^{p(n)}(G^{p(n)}_M)^2 /2qM$ and $ K^{p(n)}_{TL'}=-
 \sqrt{2}{\cal{N}}^{p(n)}G^{p(n)}_EG^{p(n)}_M$.  It should be stressed that
 the usefulness of Eqs.  (\ref{RTHe}) and (\ref{RTLHe}) in the extraction
 of $G_{E}^{n}$ and $G_{M}^{n}$ is related to the possibility that ${\cal
 H}^N_{T'(TL')}$ be independent of nucleon form factors, at least at qe
 peak.  Within PWIA, actually, the functions ${\cal H}^N_{T'(TL')}$ and
 ${\cal T}^N_{T'(TL')}$ become independent of nucleon form factors at the
 qe peak, and even almost independent of $Q^2$.  The first feature can be
 immediately seen, since at qe peak $p \approx 0$, $E_p \approx M$ and
 $\bar{\nu}=Q^2/2M$; thus retaining only $p$-leading terms in Eqs.
 (\ref{RT'}) and (\ref{RTL'}) one has for $^3$He,

\be {\cal H}^N_{T'} \left(Q^2,\nu_{peak} \right)\approx {\cal
F}^{^3He,N}_{T'}\left(Q^2,\nu_{peak} \right)= ~~~~~~~~~~~~ ~~~~~~\nonumber
\\ =2\pi~\int \limits_{E_{min}} \limits^{E_{max}(Q^2,\nu_{peak})}dE\int
\limits_{p_{min}(Q^2,\nu_{peak},E)}\limits^{p_{max}(Q^2,\nu_{peak},E)}
p~dp~ \left( B_1^{N}\!\left( p,E \right)~+B_2^{N}\!\left( p,E \right)cos^2
\alpha \right)~~~~~~~~~~~~~ \label{FTHe} \ee \be {\cal H}^N_{TL'}
\left(Q^2,\nu_{peak} \right)\approx {\cal
F}^{^3He,N}_{TL'}\left(Q^2,\nu_{peak} \right)=~~~~~~~~~~~~~~~~~~\nonumber
\\ =2\pi~\int \limits_{E_{min}} \limits^{E_{max}(Q^2,\nu_{peak})}dE\int
\limits_{p_{min}(Q^2,\nu_{peak},E)}\limits^{p_{max}(Q^2,\nu_{peak},E)} p~dp
~ \left(B_1^{N}\!\left( p,E\right)~+{B_2^{N}\!\left( p,E \right) \over 2
}sin^2 \alpha \right) ~~~~~~~~~~~~ \label{FTLHe} \ee For $^3$H the same
approximations hold, but the structure functions $B_{0,1,2}$, entering the
analogous equations, are the appropriate ones for such a nucleus, i.e.
they are obtained without Coulomb interaction and exchanging the proton
with the neutron.

\indent In Fig.  5, i) the functions ${\cal
H}^N_{T'(TL')}\left(Q^2,\nu_{peak} \right)$ and ${\cal
T}^N_{T'(TL')}\left(Q^2,\nu_{peak} \right)$ calculated by using the full
PWIA expressions (cf.  Eqs.(\ref{RT'}), (\ref{RTL'}) and (\ref{HT})) and
the Galster nucleon form factors \cite{Gal}; and ii) their approximations
$\cal F^{^3He,N}_{T'(TL')}\left(Q^2,\nu_{peak} \right)$ and $\cal
F^{^3H,N}_{T'(TL')}\left(Q^2,\nu_{peak} \right)$ (see Eqs.(\ref{FTHe}) and
(\ref{FTLHe})) are shown.  Since the approximations are quite good, this
indicates that, within PWIA, the functions ${\cal
H}^N_{T'(TL')}\left(Q^2,\nu_{peak} \right)$ and ${\cal
T}^N_{T'(TL')}\left(Q^2,\nu_{peak} \right)$ become independent of the
nucleon form factors, namely the factorization of the latter quantities out
of the nuclear structure is confirmed at a large extent.  Therefore, the
expectation of extracting information on the neutron properties from
polarized $^3$He is strengthened (cf.  \cite{BW}, where only the
spin-dependent momentum distribution was considered).  It should be pointed
out that using different nucleon form factors, such as the Gari-Krumpelmann
\cite{Gari} or the Hoehler \cite{Hoe} ones, the factorization is even
better.  Moreover, the nuclear structure functions, ${\cal
H}^N_{T'(TL')}\left(Q^2,\nu_{peak} \right)$ and ${\cal
T}^N_{T'(TL')}\left(Q^2,\nu_{peak} \right)$, have a constant behaviour over
a wide range of $Q^2$ and this striking feature can be of great help in
disentangling nucleon form factors from the nuclear structure, as shown in
Sect.  5.  Finally, if we disregard the effects of the Coulomb force (an
effect of the order of a few percent for the neutron in $^3$He) the
functions ${\cal T}^p_{T'(TL')}\left(Q^2,\nu_{peak} \right)$ measured in
$^3$H, represent a very good approximations for ${\cal
H}^n_{T'(TL')}\left(Q^2,\nu_{peak} \right)$.

\indent In what follows, after a general analysis of the responses, we will
sketch a possible way for minimizing the model dependence just exploiting
the above mentioned features of the polarized responses.

\indent In Figs.  6(a,b) and 7(a,b) our PWIA results of $R_{T'}$ and
$R_{TL'}$ for $^3$He and $^3$H are shown through the ratios $R_{T'}/G_D^2$
and $R_{TL'}/G_D^2$ ($G_D=1/(1+Q^2/0.71)^2$), at qe peak and for $Q^2$ up
to $ 2 ~(GeV/c)^2$.  The variations due to the different choices of two-
and three-body nuclear forces or to the presence of the Coulomb interaction
remain less than 5-6$\%$ over the whole range of $Q^2$ explored.  It turns
out that for $^3$He the effects due to Coulomb potential and TNI are
larger, in percentage, in the proton contribution than in the neutron one
(cf.  also the values of the effective polarization listed in Table II,
where the same behaviour can be found).  The Coulomb potential negligibly
affects $R^{^3He}_{T'}$, since it yields a different sign effect in the
proton and neutron contributions.

 In view of the analysis we have carried out, Figs.  6(c) and 7(c) are of
particular interest, since they illustrate the relevance of the proton
contribution in $^3$He and the tiny effect of the neutron in $^3$H.  We can
see that while the neutron can be safely disregarded in $^3$H (the relative
contribution is less than 3 $\%$), this fact does not occur for the proton
in $^3$He.  In the kinematical interval we have considered, for
$R_{TL'}^{^3He}$ the relative proton contribution ranges between 80$\%$, at
low values of $Q^2$, and 40$\%$, at the highest ones.  For $R_{T'}^{^3He}$
the proton contribution is not as dramatic as in the case of
$R_{TL'}^{^3He}$.  As a consequence of this limited effect $R_{T'}^{^3He}$
and $R_{T'}^{^3H}$ are nearly proportional, i.e.  $R_{T'}^{^3He}\approx
(\mu_n/\mu_p)^2 R_{T'}^{^3H}$, cf.  Figs.  6(a) and 7(a); the experimental
evidence of this proportionality could be an indication of the smallness of
the proton contribution.  However, even for $R_{T'}^{^3He}$ this
contribution is not negligible ($\approx~10\%$) and has a value about twice
larger than the uncertainties due to ISI.  The conclusions that can be
drawn from Figs.  6 and 7 are in order:  i) the neutron in $^3$H can be
safely disregarded at a level of a few percent; thus a measurement of
$R_{T'(TL')}^{^3H}$ allows to check PWIA predictions such as factorizable
responses, an almost constant behaviour of ${\cal
T}^N_{T'}\left(Q^2,\nu_{peak} \right)$ and ${\cal
T}^N_{TL'}\left(Q^2,\nu_{peak} \right)$, and their equality; ii) if the
small Coulomb effects are disregarded, one can apply the isospin symmetry
in order to identify ${\cal H}^n_{T'(TL')}$ with ${\cal T}^p_{T'(TL')}$,
obtained from measurements of $^3$H responses, and in this way one could
reduce the model dependence in the extraction of neutron form factors; iii)
the proton contribution in $^3$He cannot be neglected, even at high $Q^2$.
As a final remark, let us note that an experimental observation of a
constant behaviour of ${\cal T}^p_{T'(TL')}\left(Q^2,\nu_{peak} \right)$
could give relevant information both on the approximation of the nucleon
form factors in the three-nucleon system as the free ones and on the
prescriptions for the off-energy-shell em nucleon current.

In the following Section we will illustrate how to minimize the model
dependence related to the above mentioned proton contribution.

\bigskip

\noindent{\bf 5.  The extraction of $G_{E}^{n}$ and $ G_{M}^{n}$}

\bigskip

\indent The almost constant behaviour of the proton structure function
${\cal H}^p_{TL'}\left(Q^2,\nu_{peak} \right)$, shown in Fig.  5(b),
suggests a possible way for extracting information on the proton
contribution to $R^{^3He}_{TL'}\left(Q^2,\nu_{peak} \right)$.  As a matter
of fact if we divide the {\em total response}
$R^{^3He}_{TL'}\left(Q^2,\nu_{peak} \right)$ by the proton form factors,
assumed well-known, or better by $K^{p}_{TL'}\left(Q^2 \right)=-
2\sqrt{2}~G^{p}_MG^{p}_E$, we obtain, within PWIA, an almost constant term
(${\cal H}^p_{TL'}$) {\em plus} a $Q^2$-dependent one, i.e.  (see
Eq.(\ref{RTLHe})) \be {R^{^3He}_{TL'}\left(Q^2,\nu_{peak} \right)\over
K^p_{TL'}\left(Q^2,\nu_{peak} \right)}={\cal H}^p_{TL'}\left(Q^2,\nu_{peak}
\right)+{G_{E}^{n}\!\left({Q}^2\right ) G_{M}^{n}\!\left({Q}^2\right )
\over 2 G_{E}^{p}\!\left({Q}^2\right ) G_{M}^{p}\!\left({Q}^2\right )}{\cal
H}^n_{TL'}\left(Q^2,\nu_{peak} \right) \label{RTKp} \ medskip \ee For small
values of the momentum transfer, the $Q^2$-dependent term should be linear
in $Q^2$, due to the presence of the neutron charge form factor, $G^n_E$,
and the almost constant behaviour of the neutron structure function ${\cal
H}^n_{TL'}\left(Q^2,\nu_{peak} \right)$; therefore Eq.(\ref{RTKp}) becomes
\be {R^{^3He}_{TL'}\left(Q^2,\nu_{peak} \right)\over
K^p_{TL'}\left(Q^2,\nu_{peak} \right)}\approx {\tilde {\cal {H}}}^p_{TL'}+
\alpha ~Q^2 \label{lin} \ee where ${\tilde {\cal{H}}}^p_{TL'}$ is a
constant value.  This behaviour is confirmed by the direct calculations
presented in Fig.  8, where the ratio ${R^{^3He}_{TL'}\left(Q^2,\nu_{peak}
\right)/ K^{p}_{TL'}\left(Q^2 \right)}$, evaluated within PWIA, is plotted
for $0.1 \le Q^2 \le 0.3~(GeV/c)^2$.  This range of momentum transfer can
be easily understood, since at very low values of $Q^2$ the response
vanishes, and therefore we have to move from this particular region, while
at high values of momentum transfer $G^n_E/G^p_E$ is no more linear in
$Q^2$ (see, e.g.  the parametrizations of \cite{Gal,Hoe}).

\indent The extraction of the proton contribution to
$R^{^3He}_{TL'}\left(Q^2,\nu_{peak} \right)$ should proceed as follows:
after measuring $R^{^3He}_{TL'}\left(Q^2,\nu_{peak} \right)$ in the
proposed range, one should check whether the data exhibit the linear
behaviour shown in Fig.  8, and in the positive case one can determine
${\tilde{\cal H}}^p_{TL'}$, to be compared with theoretical predictions.
From the experimental value of ${\tilde{\cal H}}^p_{TL'}$, one obtains
$R^{^3He, p}_{TL'}\left(Q^2,\nu_{peak} \right)$ over the whole range of
$Q^2$, and singles out the neutron contribution $R^{^3He,
n}_{TL'}\left(Q^2,\nu_{peak} \right)$, which is sensitive to the neutron
charge form factor.  It should be pointed out that the results shown in
Fig.  8 rely on the factorization of the response, as discussed in Sect.
4, and on the absence of effects such as FSI; therefore whether the linear
behavior predicted by the PWIA (cf.  Eq.(\ref{lin})) is not observed,
essential information on both the reaction mechanism and the presence of
other effects can be extracted.  For instance, a linear behaviour in $Q$,
and not in $Q^2$, would mean that the variation of
$R^{^3He}_{TL'}\left(Q^2,\nu_{peak} \right)$ is dominated by
$R^{^3He,p}_{TL'}$, since the $Q^2$-dependence of the neutron part in
$R^{^3He}_{TL'}$ is always governed by $G_E^n$.  In this case we can
compare the theoretical calculations of the proton contribution, that are
not affected by unknown form factors, with the experimental data, obtaining
constraints on $R^{^3He,p}_{TL'}$.  Therefore, in any case, either PWIA
holds or not, the measurement of $R^{^3He}_{TL'}\left(Q^2,\nu_{peak}
\right)$ at low $Q^2$ will yield valuable information on the proton, to be
used in further steps.

Once we have an experimental estimate of the proton contribution to the
transverse-longitudinal response, $R^{^3He,p}_{TL'}\left(Q^2,\nu_{peak}
\right)$, we could achieve an estimate of proton contribution to the
transverse response, $R^{^3He,p}_{T'}\left(Q^2,\nu_{peak} \right)$, by
using the $\beta$-kinematics analyzed in \cite{CPS}.  The choice of the
polarization angle $\beta$, with $cos \beta=\vec{S}_A \cdot
\vec{k}_1/|\vec{k}_1|$ is suggested by:  i) the direct connection with the
experimental set-up and ii) more important, the low dependence of
$\beta_{critic}$ (see below) upon kinematical conditions.  It turns out
that the proton contribution to the polarized cross section,
$\Delta^p\left(Q^2,\nu_{peak} \right)$, (cf.  Eq.  (\ref{Delta})) vanishes
when the polarization angle $\theta^*$, reaches a critical value or, see
Eq.(\ref{ang}), when the polarization angle $\beta$ reaches
$\beta_{critic}$.  An estimate of such an angle, at the qe peak, can be
obtained by measuring the protons knocked out along the direction of
$\vec{q}$, since at the qe peak protons should be emitted preferably along
such a direction.  It is worth noting that an exclusive measurement of
protons is much easier than the one of neutrons, and moreover row data are
sufficient for estimating $\beta_{critic}$, since one has only to determine
the polarization angle where the polarized response
$\Delta^p\left(Q^2,\nu_{peak} \right)$ changes sign.  After determining
$\beta_{critic}$ at the qe peak, one can obtain the corresponding
$\theta^*$, see Eq.(\ref {ang}), and finally
$R^{^3He,p}_{T'}\left(Q^2,\nu_{peak}
\right)/R^{^3He,p}_{TL'}\left(Q^2,\nu_{peak} \right)$ from the equation
$\Delta^p\left(Q^2,\nu_{peak} \right)=0$, see Eq.(\ref {Delta}).  It should
be pointed out that the measurement of $\beta_{critic}$ is related to a
ratio of response functions, and therefore, even in presence of other
effects, the PWIA prediction should represent a good approximation;
furthermore, it turns out that $\beta_{critic}$ does not sizably vary as
function of $Q^2$ (at most a few percent), while $\theta^*_{critic}$ does.
In Fig.  9, the angle $\beta_{critic}$ is shown for different ISI, at a
scattering angle $\theta_e=50^o$.  It should be pointed out that an error
of $\pm1^o$ on the measurement of $\beta_{critic}$ produces an uncertainty
of the order 5$\%$ on the ratio $R^{^3He,p}_{T'}/R^{^3He,p}_{TL'}$ while an
error of $\pm2^o$ produces an uncertainty of the order 12$\%$.  Even a
large uncertainty on $R^{^3He,p}_{T'}$ does not prevent the extraction of
the neutron magnetic form factor, since $R^{^3He,p}_{T'}$ is small, of the
order of 10$\%$ of the total response, as shown in Fig.6(c).

 In conclusion, through an estimate of $R^{^3He,p}_{TL'}$, from the low
$Q^2$ behaviour of the total response $R^{^3He}_{TL'}$, and an estimate of
$R^{^3He,p}_{T'}$, from $\beta_{critic}$, one can obtain the neutron
contribution to the total responses $R^{^3He}_{TL'}$ and $R^{^3He}_{T'}$,
respectively.  From the ratio $R^{^3He,n}_{TL'}/R^{^3He,n}_{T'}$ one can
obtain the ratio $G_{E}^{n}/ G_{M}^{n}$, assuming that the functions
${\cal{H}}^n_{TL'}$ and ${\cal{H}}^n_{T'}$ are equal, as in the case of
PWIA; moreover, introducing a theoretical prediction for
${\cal{H}}^n_{T'(TL')}$ one could extract $G_{E}^{n}$ and $ G_{M}^{n}$
separately.

 Finally a measurement of the polarized $^3$H could give the possibility of
an almost model-independent extraction of both the neutron form factors,
since the structure functions ${\cal{H}}^n_{TL'}$ and ${\cal{H}}^n_{T'}$
could be estimated through ${\cal{T}}^p_{TL'}$ and ${\cal{T}}^p_{T'}$,
disregarding the Coulomb effects.

\bigskip

\noindent{\bf 6.  Conclusion}

\bigskip

\indent In this paper we have presented the results of our investigation on
the em inclusive responses of polarized $^3$He and $^3$H, within the PWIA.
We have calculated the spin-dependent spectral functions \cite{CPS} of the
three-nucleon system, from bound-state wave functions, obtained using the
pair correlated hyperspherical-harmonic expansion \cite{KVR}.  Different
realistic two- and three-body nuclear forces, such as i) the Argonne v14
potential and the RSCv8 one, and ii) the Brazil and Tucson-Melbourne
three-nucleon interactions, have been considered; moreover, the Coulomb
interaction has been taken into account in the case of $^3$He.  Then we
have evaluated the transverse-longitudinal polarized response and the
transverse one, focusing at the qe peak, for $Q^2$ up to $2~(GeV/c)^2$.
The detailed analysis of the $Q^2$-behaviour of the inclusive responses has
allowed:  i) to investigate the model dependence upon the initial state
interaction and ii) to suggest possible experiments for determining the
proton contribution to $R^{^3He}_{TL'}\left(Q^2,\nu_{peak} \right)$ and
$R^{^3He}_{T'}\left(Q^2,\nu_{peak} \right)$, which represents one of the
major obstacle in the experimental extraction of the neutron form factors.
The model dependence upon two- and three-nucleon interactions and Coulomb
potential as well, amounts to a few percent; for $^3$He the Coulomb and TNI
effects are more relevant for the proton contribution than the neutron one,
as in the case of the effective polarization.

The presence of the proton affects quite differently the polarized
responses, since in $R^{^3He}_{T'}\left(Q^2,\nu_{peak} \right)$ it is
$\approx~10~\%$, while in $R^{^3He}_{TL'}\left(Q^2,\nu_{peak} \right)$ it
ranges between 80$\%$, at low values of $Q^2$, and 40$\%$, at the highest
ones.  This proton predominance can be turned in our advantage, since PWIA
predicts a linear behaviour in $Q^2$ for the ratio
$R^{^3He}_{TL'}\left(Q^2,\nu_{peak} \right)/\left(-
2\sqrt{2}~G^{p}_M\left(Q^2 \right)G^{p}_E\left(Q^2 \right)\right)$ at low
momentum transfer, $0.1 \le Q^2 \le 0.3~ (GeV/c)^2$.  Therefore a
comparison with the experimental data can yield $R^{^3He,p}_{TL'}$ or, at
least, definite information on the proton contribution, e.g.  on the
presence of factorization and/or FSI.

The proton contribution to $R^{^3He}_{T'}$ can be measured from the ratio
$R^{^3He,p}_{T'}\left(Q^2,\nu_{peak} \right)/$
$R^{^3He,p}_{TL'}\left(Q^2,\nu_{peak} \right)$ obtained through an accurate
determination of the polarization angle $\beta_{critic}$, where the proton
contribution to the polarized cross section, see Eq.(\ref{Delta}), is
vanishing.  The PWIA prediction of this angle could be used as a reliable
guideline for the experimental measurements, since $\beta_{critic}$ depends
upon the ratio of responses, possibly less sensitive to various effects,
such as FSI, than each response separately.

The proposed measurements could allow the extraction of the neutron
contribution to the total responses and therefore an estimate of the ratio
$G_{E}^{n}/ G_{M}^{n}$.  If one introduces theoretical calculations of the
nuclear structure functions ${\cal{H}}^n_{T'(TL')}$, one could even obtain
$G_{E}^{n}$ and $G_{M}^{n}$ separately.  Finally, it should be pointed out
that a measurement of the em inclusive responses of polarized $^3$H could
give the possibility to check more directly the reaction mechanism, namely
the factorization at the qe peak (essential for extracting the neutron form
factors), and to obtain the nuclear structure functions,
${\cal{T}}^p_{T'(TL')}$.  In this case, one can substantially lower the
model dependence in the extraction of $G_{E}^n$ and $G_{M}^n$ by using the
$^3$H structure functions as an experimental estimate of the corresponding
quantities for $^3$He.

Calculations for taking into account FSI are in progress.

\bigskip

\noindent{\bf 6.  Acknowledgment}

\bigskip

\noindent The authors are very grateful to C.  Ciofi degli Atti and S.
Rosati for the encouragement in carrying on this work and for valuable
discussions.  Two of us (E.P.  and G.S.)  also acknowledge stimulating
discussions with H.  Gao and O.  Hansen.

\newpage \begin{center} FIGURE CAPTIONS \end{center}

\noindent Fig.  1(a).  The proton unpolarized spectral function of $^3$He,
obtained by using the Argonne v14 NN potential \cite{Arg} and the Coulomb
interaction (see text), vs.  the removal energy E and the nucleon momentum
$p~\equiv~|\vec{p}|$.  The deuteron channel is not presented.  \bigskip

\noindent Fig.  1(b).  The function $\mid B_{1,{1 \over 2}}^p \mid$,
obtained by using the Argonne v14 NN potential \cite{Arg} and the Coulomb
interaction (see text), vs.  the removal energy E and the nucleon momentum
$p~\equiv~|\vec{p}|$.  \bigskip

\noindent Fig.  1(c).  The function $\mid B_{2,{1 \over 2}}^p \mid$,
obtained by using the Argonne v14 NN potential \cite{Arg} and the Coulomb
interaction (see text), vs.  the removal energy E and the nucleon momentum
$p~\equiv~|\vec{p}|$.  \bigskip

\noindent Fig.  2(a).  The same as Fig.  1(a), but for the neutron.
\bigskip

\noindent Fig.  2(b).  The same as Fig.  1(b), but for the neutron.
\bigskip

\noindent Fig.  2(c).  The same as Fig.  1(c), but for the neutron.
\bigskip

\noindent Fig.  3(a).  The asymmetry of $^3$He corresponding to
 $\epsilon_1$ = 370~$MeV$, $\theta_e = -91.4^o$ and $\beta=42.5^o$, vs.
 the energy transfer $\nu$, calculated with different two-body
 interactions.  Solid line:  Av14 + Coulomb interaction; dashed line:
 RSCv8 + Coulomb interaction; dotted line:  Paris potential without Coulomb
 interaction, after \cite{Sauer}.  Dot-dashed line:  the neutron
 contribution for the Av14 + Coulomb interaction case.  The nucleon form
 factors of Ref.  \cite{Gal} have been used.  The arrow indicates the
 position of the qe peak, where one has $\theta^*\simeq 8.9^o$ and
 $Q^2=0.19 ~(GeV/c)^2$.  The experimental data are from Ref.  \cite{Gao}.
 \bigskip

\noindent Fig.  3(b).  The same as Fig.  3(a), but for different three-body
interactions.  Solid line:  Av14 + Coulomb interaction; dashed line:  the
Brazil three-body force \cite{BR} has been added to Av14 + Coulomb
interaction; dotted line:  the same as the dashed line, but for the
Tucson-Melbourne three-body force \cite{TM}.  The three curves largely
overlap for both the total asymmetry and the neutron contribution.

\bigskip

\noindent Fig.  4(a).  The same as Fig.  3(a), but for $\theta_e = 70.1^o$.
The experimental data are from Ref.  \cite{Han}.  The arrow indicates the
position of the qe peak, where one has $\theta^*\simeq 87^o$ and $Q^2= .14
~(GeV/c)^2$.  \bigskip

\noindent Fig.  4(b).  The same as Fig.  3(b), but for $\theta_e~ =
70.1^o$.

\bigskip

\noindent Fig.  5(a).  The functions ${\cal
H}^n_{T'(TL')}\left(Q^2,\nu_{peak} \right)$ for the neutron in $^3$He (see
Eq.  (\ref{HT})) vs $Q^2$, and the corresponding approximation $\cal
F^{^3He,n}_{T'(TL')}\left(Q^2,\nu_{peak} \right)$ (see Eqs.  (\ref{FTHe})
and (\ref{FTLHe})).  All the curves have been obtained from the Av14 +
Coulomb interaction and the nucleon form factors of Ref.  \cite{Gal}.
Solid line:  ${\cal H}^n_{T'}\left(Q^2,\nu_{peak} \right)$ ; dot-dashed
line:  ${\cal H}^n_{TL'}\left(Q^2,\nu_{peak} \right)$; dashed line:  $\cal
F^{^3He,n}_{T'}$ (see Eq.(\ref{FTHe})); dotted line:  $\cal
F^{^3He,n}_{TL'}$ (see Eq.(\ref{FTLHe})).  The curves largely overlap and
can be hardly singled out.

\bigskip

\noindent Fig.  5(b).  The same as Fig.  5(a), but for proton in $^3$He.
\bigskip

\noindent Fig.  5(c).  The same as Fig.  5(a), but for proton in $^3$H.
\bigskip

\noindent Fig.  5(d).  The same as Fig.  5(a), but for neutron in $^3$H.

\bigskip

\noindent Fig.  6(a).  The ratio $R^{^3He}_{T'}\left(Q^2,\nu_{peak}
\right)/G_D^2(Q^2)$ (see Eq.  (\ref{RT'})) vs $Q^2$ ($G_D(Q^2)=
1/(1+Q^2/0.71)^2$).  Solid line:  the response $R_{T'}$ obtained from the
Av14 + Coulomb interaction; dot-dashed line:  the same as the solid line,
but without Coulomb interaction, (this line largely overlaps with the solid
one); dotted line:  the same as the solid line, but for RSCv8 potential;
dashed line:  the three-body interaction \cite{BR,TM} has been added to
Av14 + Coulomb interaction (the Brazil \cite{BR} and the Tucson-Melbourne
\cite{TM} three-body interactions essentially yield the same results).

\bigskip

\noindent Fig.  6(b).  The same as Fig.  6(a), but for the response
$R^{^3He}_{TL'}\left(Q^2,\nu_{peak} \right)$ (see Eq.  (\ref{RTL'})).

\bigskip

\noindent Fig.  6(c).  The ratios $R_{TL'}^{^3He,p}\left(Q^2,\nu_{peak}
\right)/R^{^3He}_{TL'}\left(Q^2,\nu_{peak} \right)$ and
$R_{T'}^{^3He,p}\left(Q^2,\nu_{peak}
\right)/R^{^3He}_{T'}\left(Q^2,\nu_{peak} \right)$ vs.  $Q^2$, at the qe
peak.  The lines are the same as in Fig.  6(a).

\bigskip

\noindent Fig.  7(a).  The same as Fig.  6(a), but for $^3H$.  Solid line:
the response $R_{T'}$ obtained from the Av14 interaction; dotted line:  the
same as the solid line, but for RSCv8 potential; dashed line:  the
three-body interaction \cite{BR,TM} has been added to Av14 interaction (the
Brazil \cite{BR} and the Tucson-Melbourne \cite{TM} three-body interactions
essentially yield the same results).  \bigskip

\noindent Fig.  7(b).  The same as Fig.  7(a), but for the response
$R^{^3H}_{TL'}\left(Q^2,\nu_{peak} \right)$ (see Eq.  (\ref{RTL'})).
\bigskip

\noindent Fig.  7(c).  The same as Fig.  6(c), but for neutron in $^3H$.
The lines are the same as in Fig.  7(a).  \bigskip

\noindent Fig.  8.  The ratio $R^{^3He}_{TL'}\left(Q^2,\nu_{peak}
\right)/K^{p}_{TL'}$, with $ K^{p}_{TL'}=- 2\sqrt{2}G^{p}_EG^{p}_M$, vs.
$Q^2$ (see text) .  The solid line represents the calculation obtained by
using in Eq.(\ref{RTL'}) the spin-dependent spectral function corresponding
to Av14 + Coulomb interaction and the Galster nucleon form factors
\cite{Gal}.  Dot-dashed line:  the ratio $R^{^3He,p}_{TL'}/K^{p}_{TL'}$,
corresponding to the proton contribution, has been shown as reference line.

\bigskip

\noindent Fig.  9.  The polarization angle $\beta_{critic}$, where the
proton contribution to the polarized cross section of $^3$He, at the qe
peak, vanishes, vs.  $Q^2$ (see Eq.(\ref{Delta})).  Solid line:  Av14 +
Coulomb interaction; dashed line:  Av14 + Coulomb interaction + three-body
forces; dotted line:  RSCv8 + Coulomb interaction (the Brazil \cite{BR} and
the Tucson-Melbourne \cite{TM} three-body interactions yield essentially
the same results).  The nucleon form factors \cite{Gal} have been used.

\newpage \begin{figure}
\psfig{figure=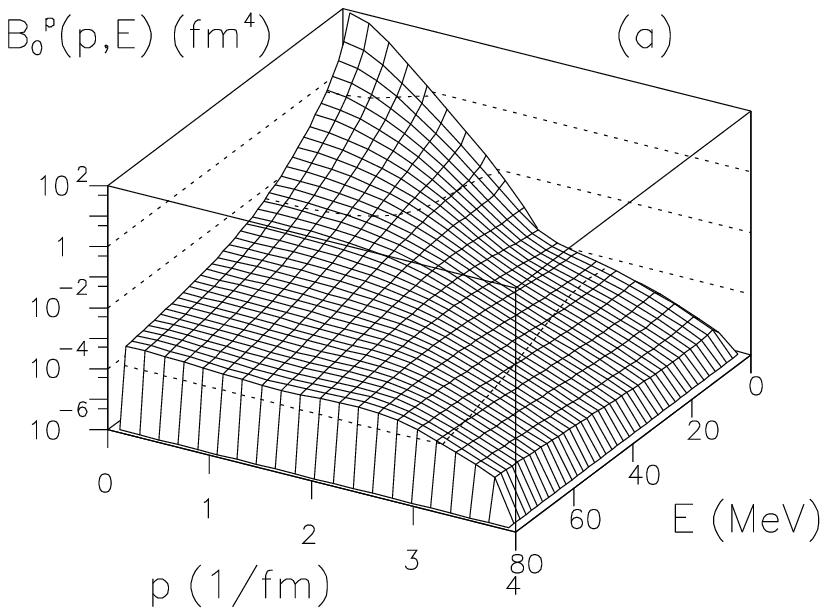,bbllx=30mm,bblly=220mm,bburx=0mm,bbury=270mm}
\psfig{figure=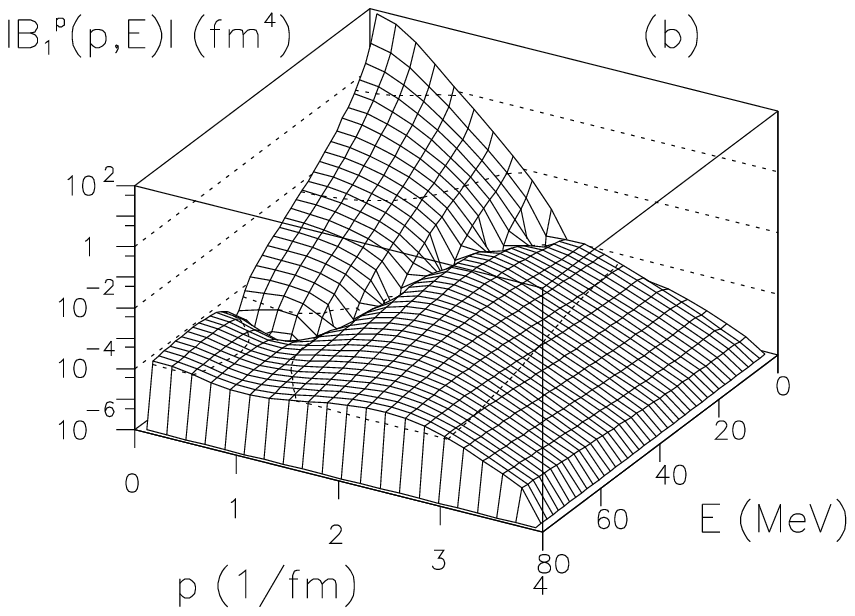,bbllx=30mm,bblly=220mm,bburx=0mm,bbury=290mm}
\psfig{figure=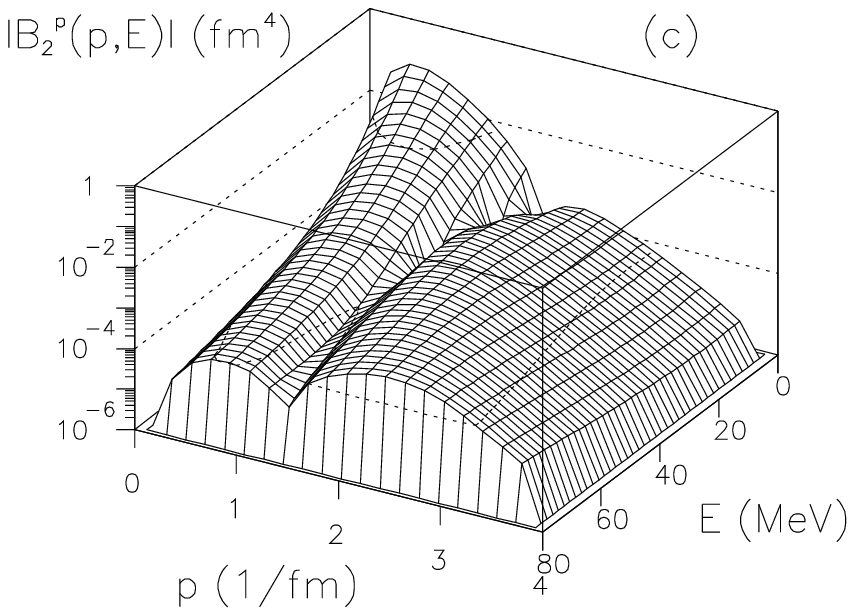,bbllx=30mm,bblly=200mm,bburx=0mm,bbury=290mm} Fig.
1 - A.  KIEVSKY, M.  VIVIANI, E.  PACE and G.  SALME' \end{figure} \newpage
\begin{figure}
\psfig{figure=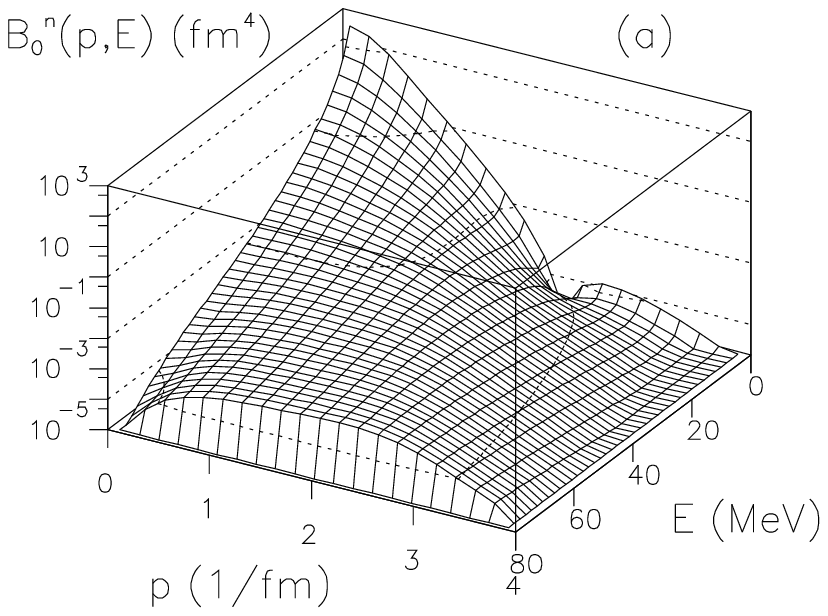,bbllx=30mm,bblly=220mm,bburx=0mm,bbury=270mm}
\psfig{figure=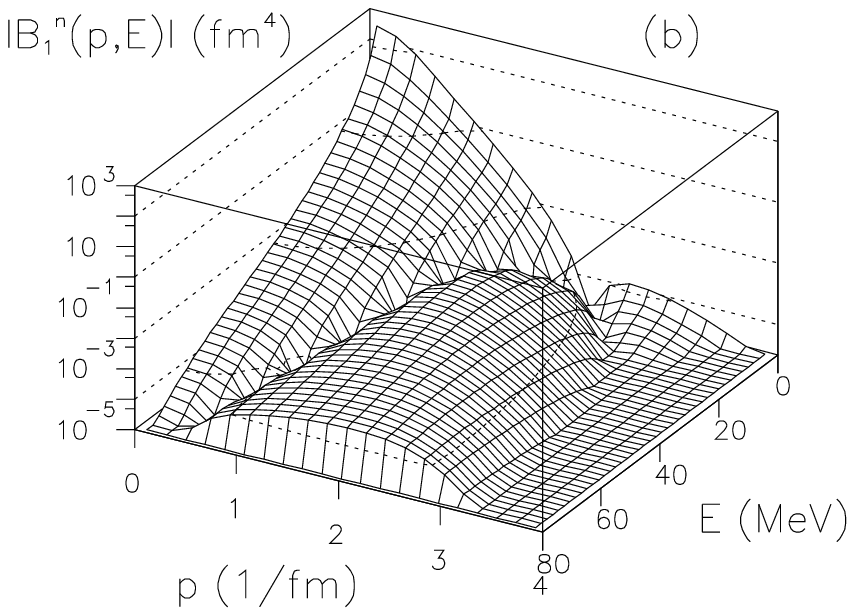,bbllx=30mm,bblly=220mm,bburx=0mm,bbury=290mm}
\psfig{figure=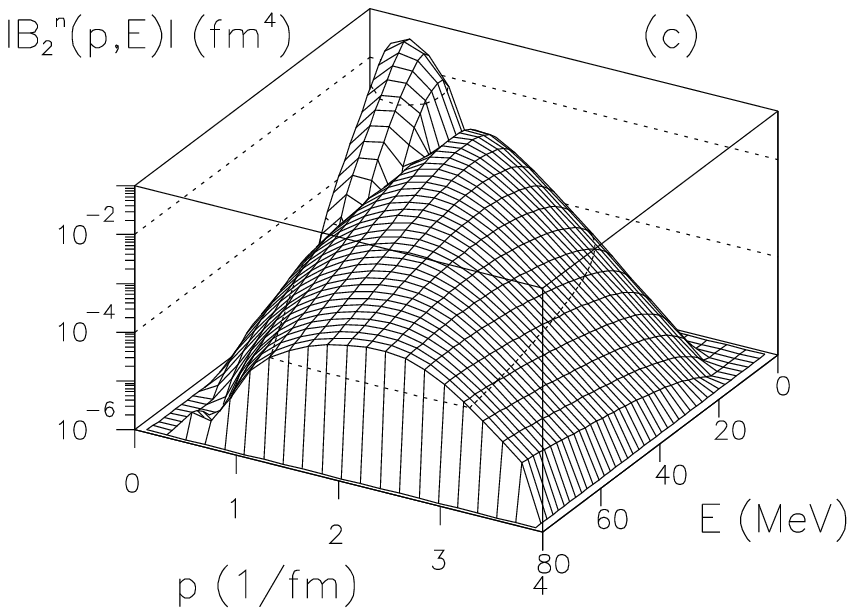,bbllx=30mm,bblly=200mm,bburx=0mm,bbury=290mm} Fig.
2 - A.  KIEVSKY, M.  VIVIANI, E.  PACE and G.  SALME' \end{figure}

\newpage \begin{figure}
\psfig{figure=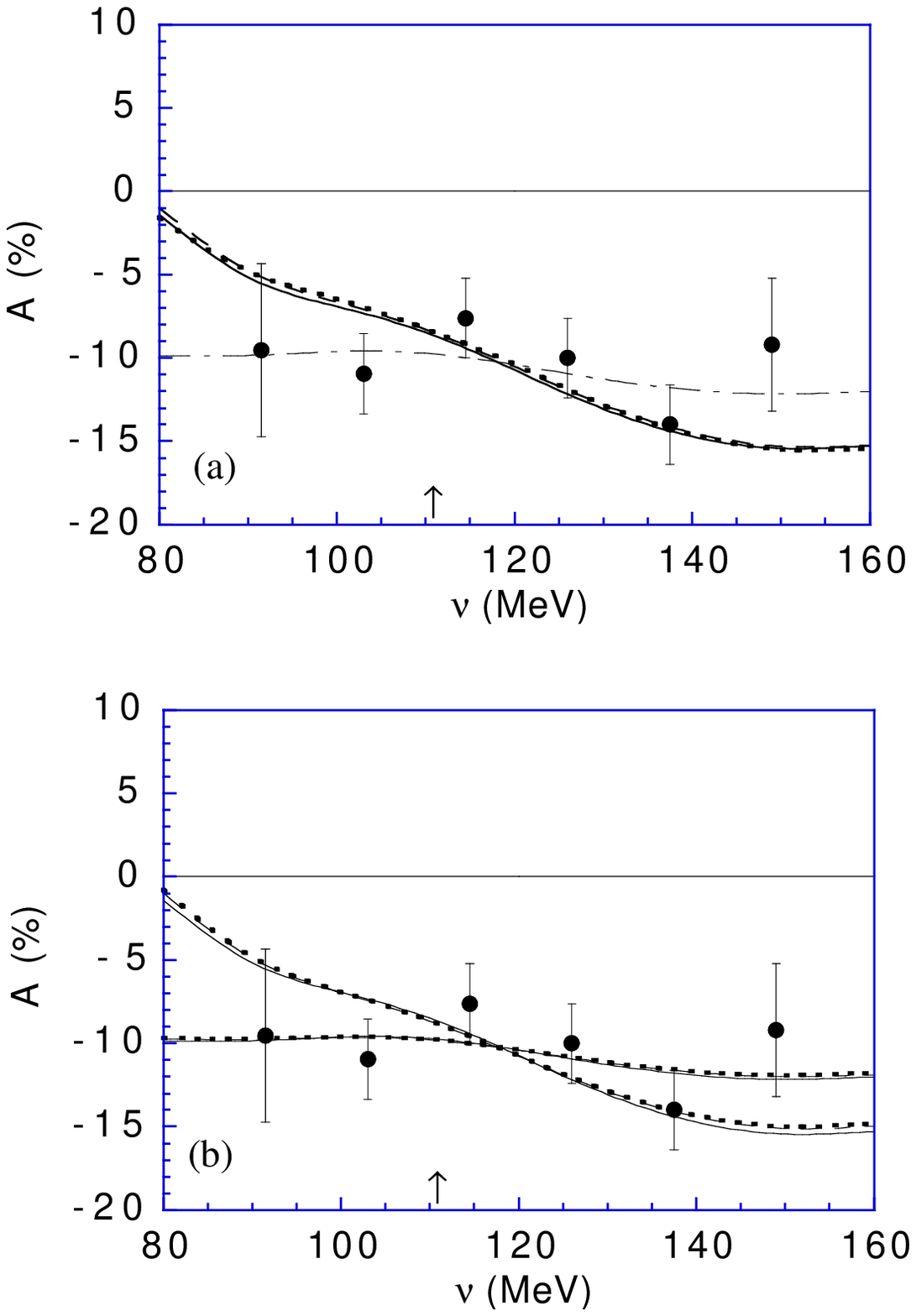,bbllx=10mm,bblly=70mm,bburx=0mm,bbury=250mm} Fig.  3 -
A.  KIEVSKY, M.  VIVIANI, E.  PACE and G.  SALME' \end{figure}

\newpage \begin{figure}
\psfig{figure=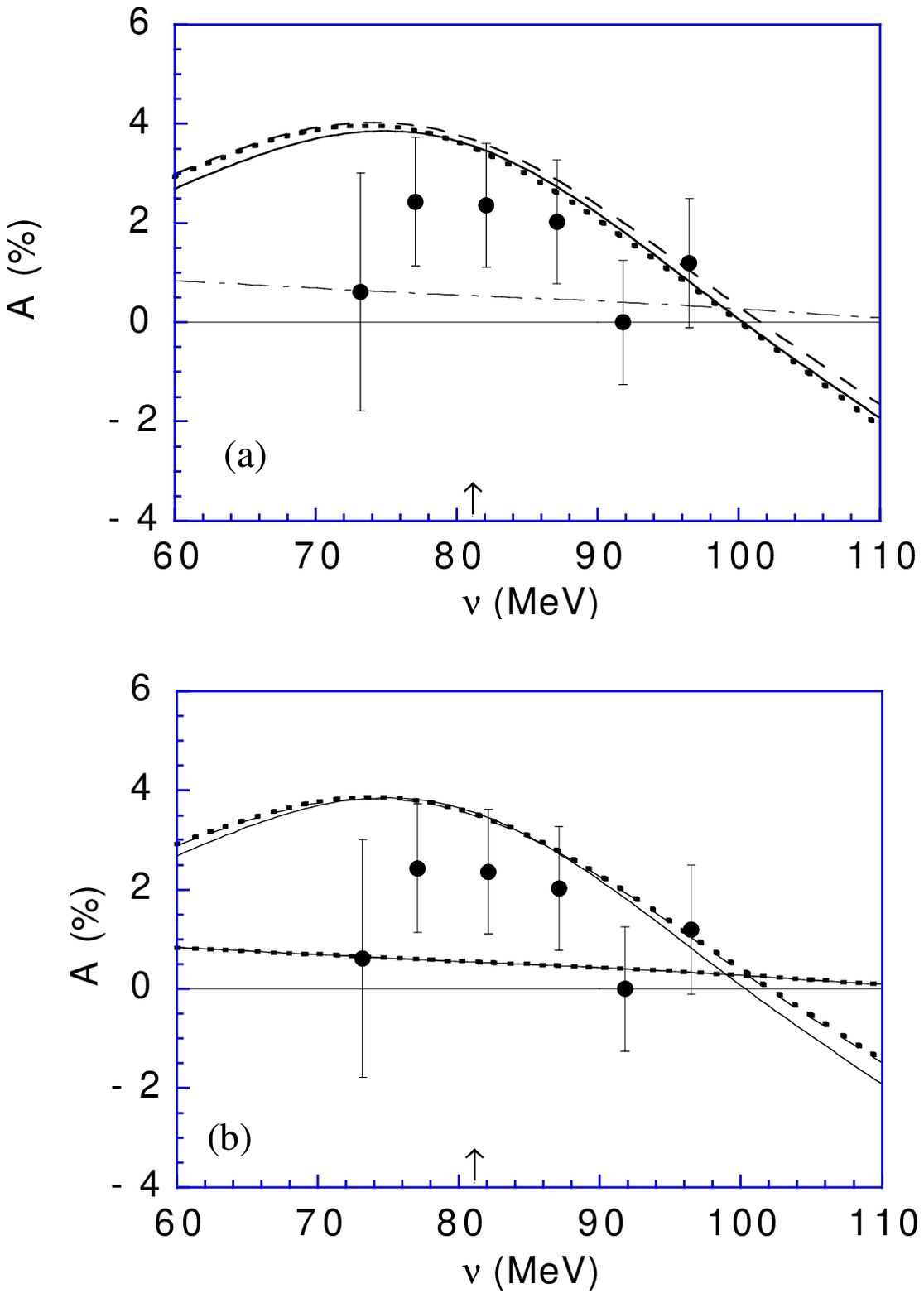,bbllx=10mm,bblly=70mm,bburx=0mm,bbury=250mm} Fig.
4 - A.  KIEVSKY, M.  VIVIANI, E.  PACE and G.  SALME' \end{figure}

\newpage \begin{figure}
\psfig{figure=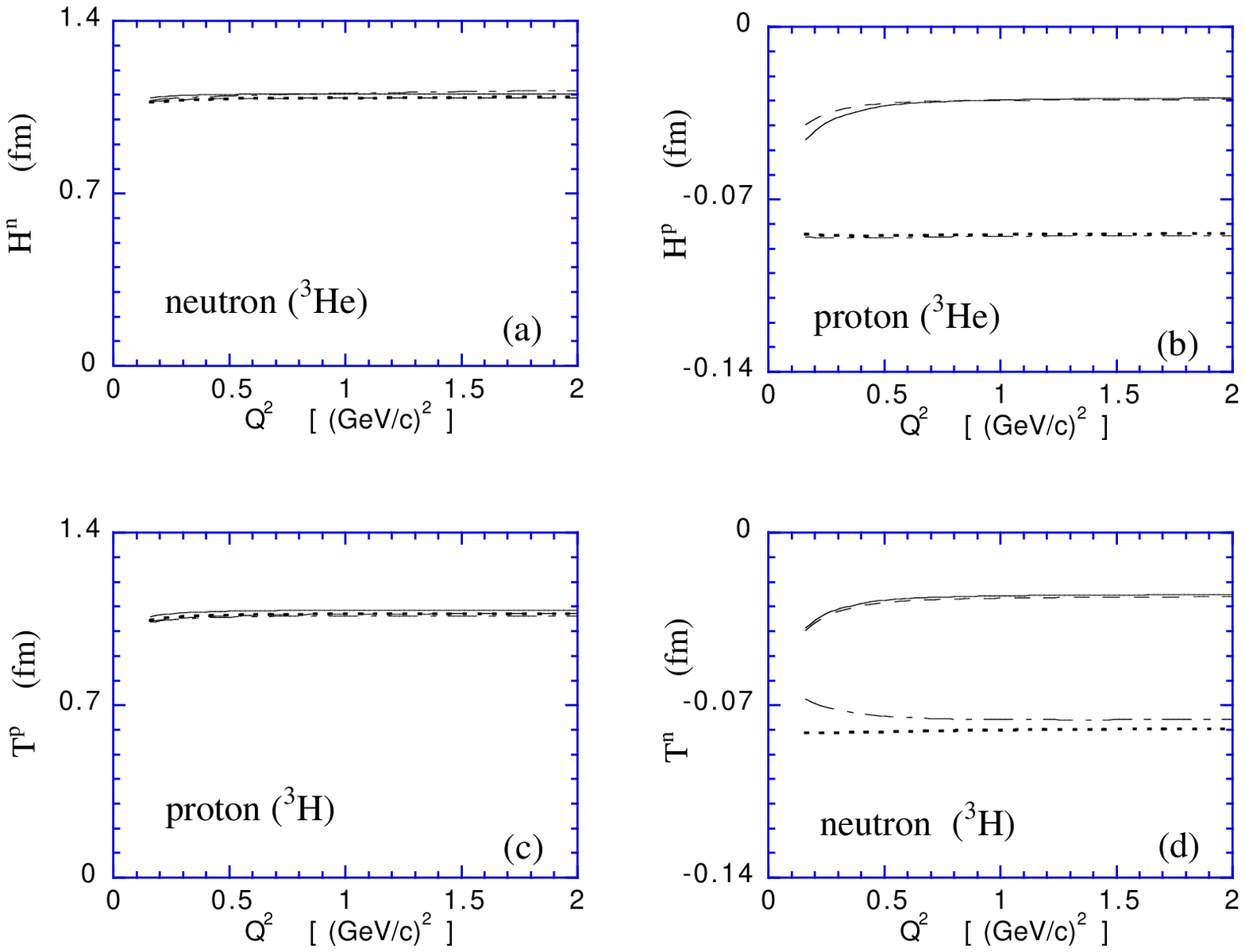,bbllx=10mm,bblly=70mm,bburx=0mm,bbury=270mm} Fig.  5 -
A.  KIEVSKY, M.  VIVIANI, E.  PACE and G.  SALME' \end{figure}

\newpage \begin{figure}
\psfig{figure=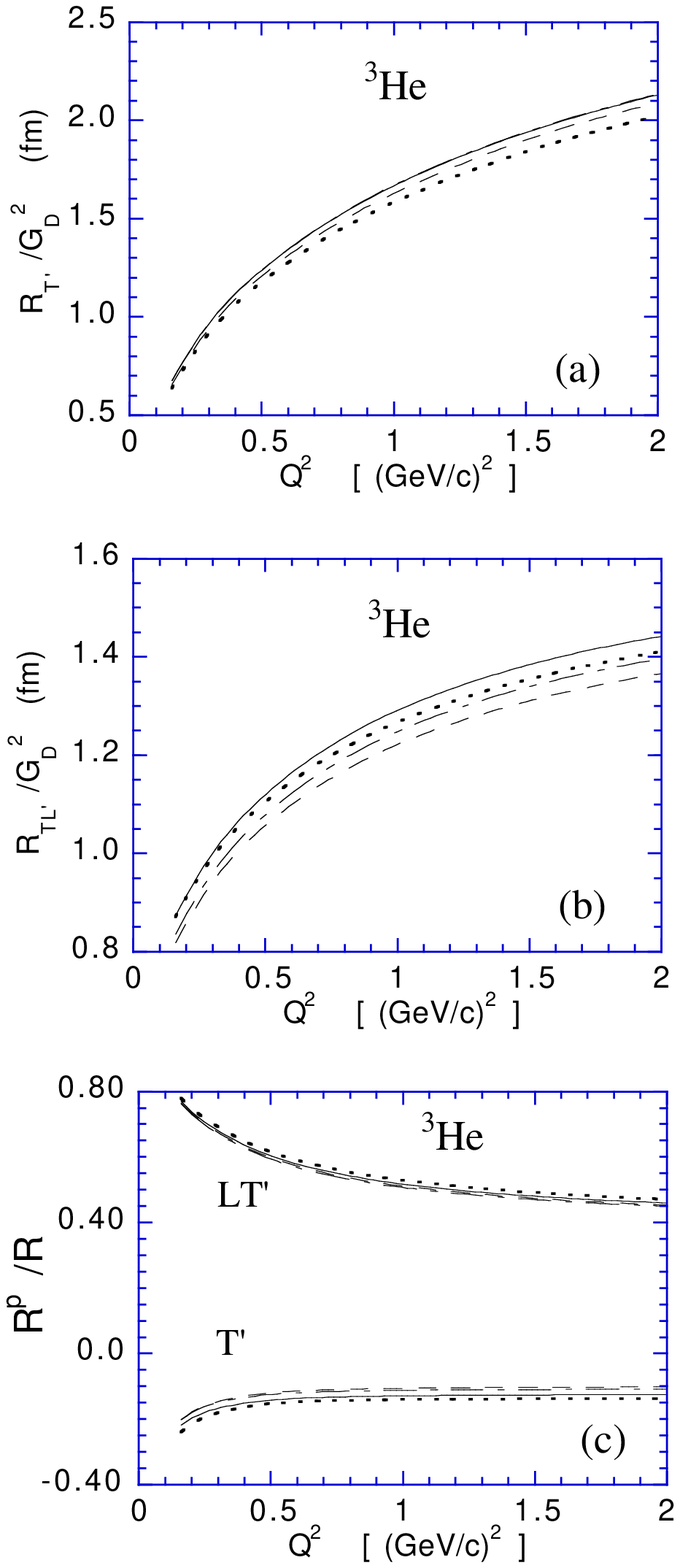,bbllx=10mm,bblly=50mm,bburx=0mm,bbury=265mm} Fig.  6
- A.  KIEVSKY, M.  VIVIANI, E.  PACE and G.  SALME' \end{figure}

\newpage \begin{figure}
\psfig{figure=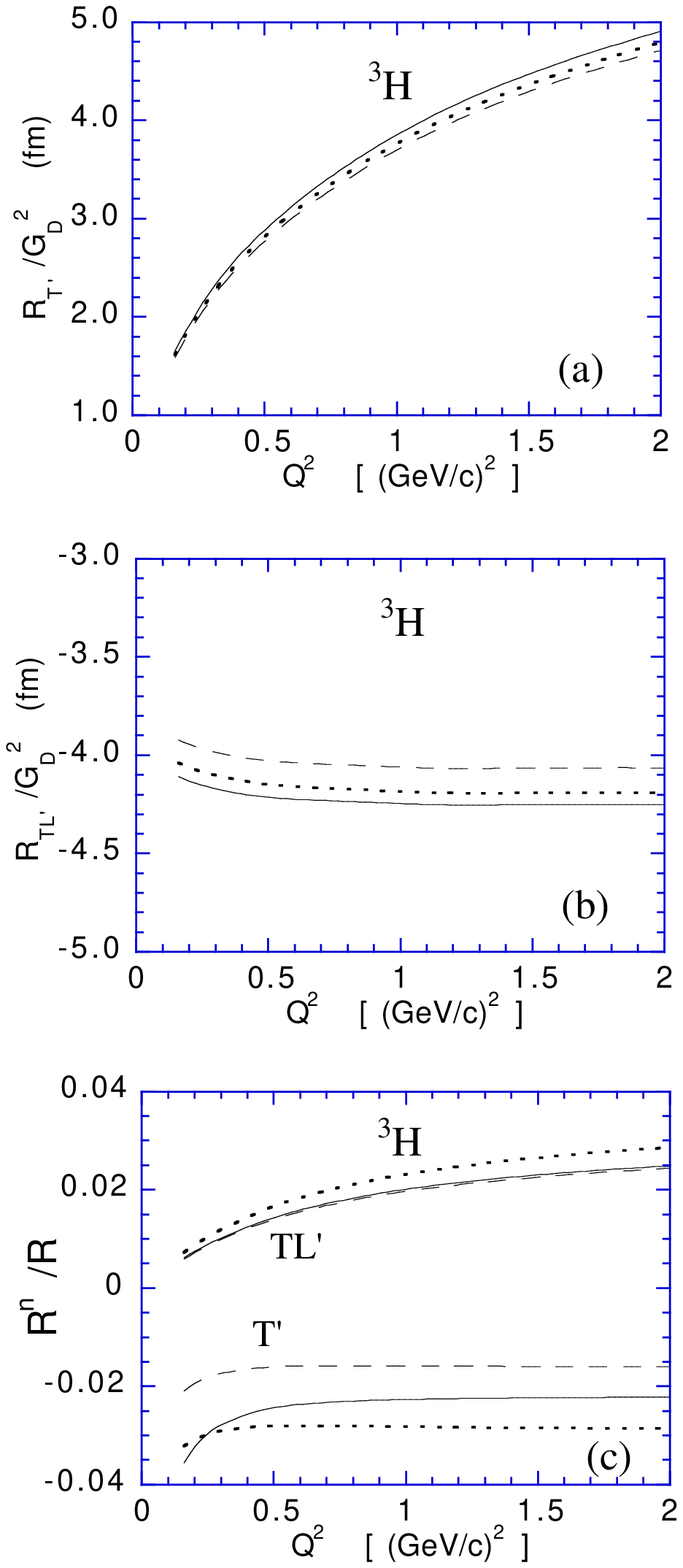,bbllx=10mm,bblly=50mm,bburx=0mm,bbury=265mm} Fig.  7
- A.  KIEVSKY, M.  VIVIANI, E.  PACE and G.  SALME' \end{figure}


\newpage \begin{figure}
\psfig{figure=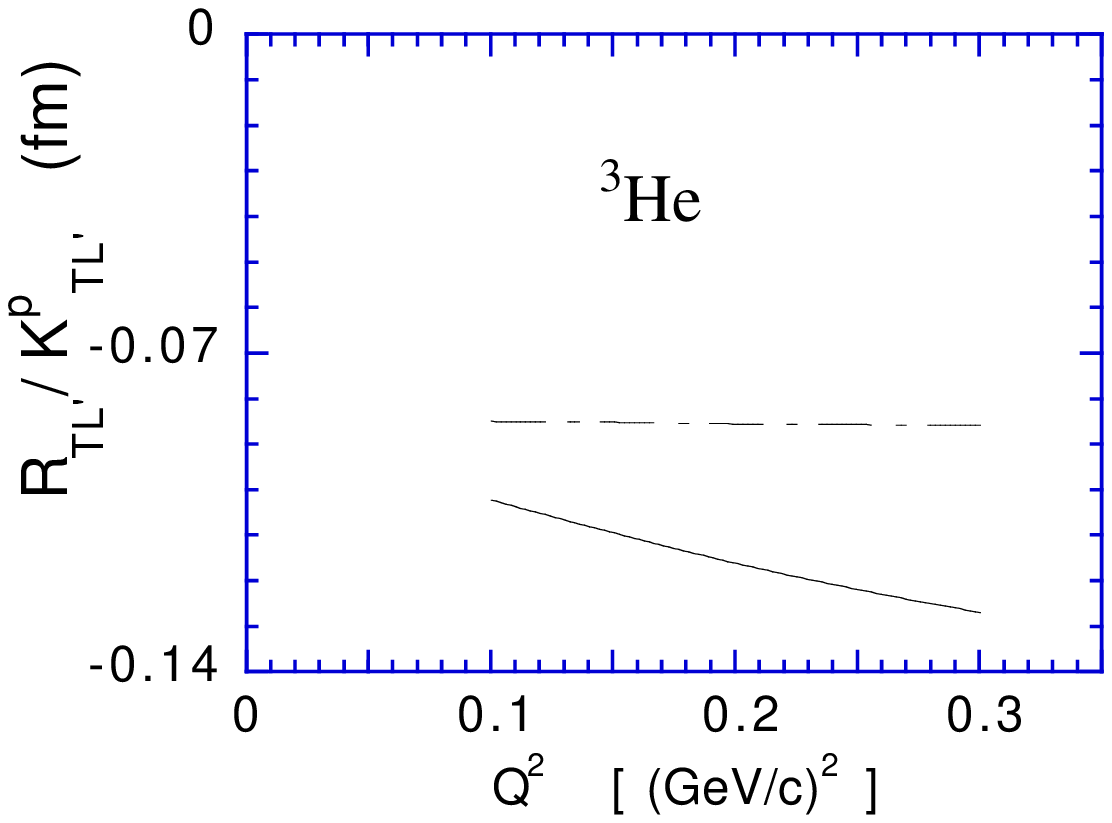,bbllx=10mm,bblly=185mm,bburx=0mm,bbury=265mm} Fig.
8 - A.  KIEVSKY, M.  VIVIANI, E.  PACE and G.  SALME' \end{figure}

\begin{figure}
\psfig{figure=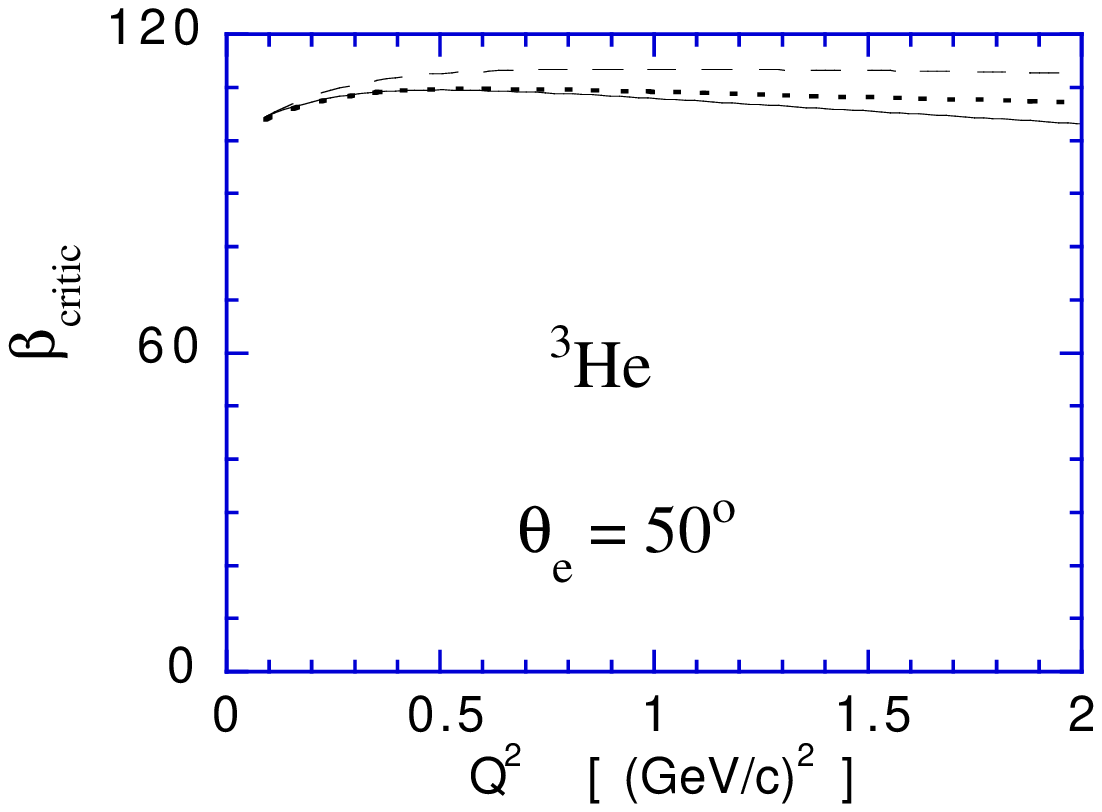,bbllx=10mm,bblly=180mm,bburx=0mm,bbury=285mm} Fig.
9 - A.  KIEVSKY, M.  VIVIANI, E.  PACE and G.  SALME' \end{figure}

\end{document}